\documentclass[
    aps,
    prd,
    twocolumn,
    superscriptaddress,
    longbibliography,
    nobibnotes
]{revtex4-1} 

\usepackage{bbm}

\usepackage[document]{}
\usepackage{microtype}
\usepackage{amsmath}
\usepackage{enumitem}
\usepackage{bbold}
\usepackage{amsfonts}
\usepackage{amssymb}
\usepackage{graphicx}
\usepackage{color}
\usepackage{accents}
\usepackage{caption}
\usepackage{dcolumn}
\usepackage[normalem]{ulem}
\usepackage[colorlinks=true, citecolor=blue, urlcolor=blue,linkcolor=blue]{hyperref}
\usepackage{enumitem}
\usepackage{dsfont}
\usepackage{makecell}
\usepackage{multirow}
\usepackage{mathtools}
\usepackage{upgreek}
\usepackage{float}
\usepackage{adjustbox}
\usepackage{siunitx}
\usepackage{pgfplots}
\pgfplotsset{compat=1.18}
\usepgfplotslibrary{groupplots,units}

\DeclareMathOperator{\Tr}{Tr}

\DeclarePairedDelimiter{\bra}{\langle}{\rvert}
\DeclarePairedDelimiter{\ket}{\lvert}{\rangle}
\DeclarePairedDelimiterX{\braket}[2]{\langle}{\rangle}{#1\delimsize{\vert}\mathopen{#2}}
\DeclarePairedDelimiter{\abs}{\lvert}{\rvert}

\DeclarePairedDelimiter{\expect}{\langle}{\rangle}

\newcommand{\estabserr}{\varDelta}
\newcommand{\op}[1]{\hat{#1}}

\usepackage{tikz}
\usetikzlibrary{quantikz}

\makeatletter 
\newsavebox{\@brx}
\newcommand{\llangle}[1][]{\savebox{\@brx}{\(\m@th{#1\langle}\)}%
  \mathopen{\copy\@brx\kern-0.5\wd\@brx\usebox{\@brx}}}
\newcommand{\rrangle}[1][]{\savebox{\@brx}{\(\m@th{#1\rangle}\)}%
  \mathclose{\copy\@brx\kern-0.5\wd\@brx\usebox{\@brx}}}
\makeatother

\newlength{\dhatheight} 

\newcommand{\qed}{\nobreak \ifvmode \relax \else
      \ifdim\lastskip<1.5em \hskip-\lastskip
      \hskip1.5em plus0em minus0.5em \fi \nobreak
      \vrule height0.75em width0.5em depth0.25em\fi}

\makeatletter
\def\@ssect@ltx#1#2#3#4#5#6[#7]#8{%
  \def\H@svsec{\phantomsection}%
  \@tempskipa #5\relax
  \@ifdim{\@tempskipa>\z@}{%
    \begingroup
      \interlinepenalty \@M
      #6{%
       \@ifundefined{@hangfroms@#1}{\@hang@froms}{\csname @hangfroms@#1\endcsname}%
       {\hskip#3\relax\H@svsec}{#8}%
      }%
      \@@par
    \endgroup
    \@ifundefined{#1smark}{\@gobble}{\csname #1smark\endcsname}{#7}%
  }{%
    \def\@svsechd{%
      #6{%
       \@ifundefined{@runin@tos@#1}{\@runin@tos}{\csname @runin@tos@#1\endcsname}%
       {\hskip#3\relax\H@svsec}{#8}%
      }%
      \@ifundefined{#1smark}{\@gobble}{\csname #1smark\endcsname}{#7}%
      \addcontentsline{toc}{#1}{\protect\numberline{}#8}%
    }%
  }%
  \@xsect{#5}%
}%
\makeatother

\begin{document}

\title{Dynamical simulations of many-body quantum chaos on a quantum computer}

\author{Laurin E. Fischer}\altaffiliation{These authors contributed equally to this work}
\affiliation{IBM Quantum, IBM Research Europe -- Zurich, 8803 R\"{u}schlikon, Switzerland}
\affiliation{Theory and Simulation of Materials, {\'E}cole Polytechnique F{\'e}d{\'e}rale de Lausanne, 1015 Lausanne, Switzerland}

\author{Matea Leahy}\altaffiliation{These authors contributed equally to this work}
\affiliation{Algorithmiq Ltd, Kanavakatu 3C, 00160 Helsinki, Finland}

\author{Andrew Eddins}
\affiliation{IBM Quantum, IBM Research - Cambridge, Cambridge, MA 02142, USA}

\author{Nathan Keenan}
\affiliation{IBM Quantum, IBM Research Europe - Dublin, IBM Technology Campus, Dublin 15, Ireland}
\affiliation{School of Physics, Trinity College Dublin, College Green, Dublin 2, D02K8N4, Ireland}

\author{Davide Ferracin}
\affiliation{Algorithmiq Ltd, Kanavakatu 3C, 00160 Helsinki, Finland}
\affiliation{Dipartimento di Fisica “Aldo Pontremoli”, Università degli Studi di Milano, Via Giovanni Celoria 16, 20133 Milano, Italy}

\author{Matteo A. C. Rossi}
\affiliation{Algorithmiq Ltd, Kanavakatu 3C, 00160 Helsinki, Finland}

\author{Youngseok Kim}
\affiliation{IBM Quantum, IBM T.J. Watson Research Center, Yorktown Heights, NY 10598, USA}

\author{Andre He}
\affiliation{IBM Quantum, IBM T.J. Watson Research Center, Yorktown Heights, NY 10598, USA}

\author{Francesca Pietracaprina}
\affiliation{Algorithmiq Ltd, Kanavakatu 3C, 00160 Helsinki, Finland}

\author{Boris Sokolov}
\affiliation{Algorithmiq Ltd, Kanavakatu 3C, 00160 Helsinki, Finland}

\author{Shane Dooley}
\affiliation{School of Physics, Trinity College Dublin, College Green, Dublin 2, D02K8N4, Ireland}

\author{Zoltán Zimborás}
\affiliation{Algorithmiq Ltd, Kanavakatu 3C, 00160 Helsinki, Finland}
\affiliation{HUN-REN Wigner RCP,  P.O. Box 49 Budapest, Hungary}

\author{Francesco Tacchino}
\affiliation{IBM Quantum, IBM Research Europe -- Zurich, 8803 R\"{u}schlikon, Switzerland}

\author{Sabrina Maniscalco}
\affiliation{Algorithmiq Ltd, Kanavakatu 3C, 00160 Helsinki, Finland}

\author{John Goold}  \email{gooldj@tcd.ie}
\affiliation{School of Physics, Trinity College Dublin, College Green, Dublin 2, D02K8N4, Ireland}
\affiliation{Trinity Quantum Alliance, Unit 16, Trinity Technology and Enterprise Centre, Pearse Street, Dublin 2, D02YN67, Ireland}
\affiliation{Algorithmiq Ltd, Kanavakatu 3C, 00160 Helsinki, Finland}

\author{Guillermo García-Pérez}  \email{guille@algorithmiq.fi}
\affiliation{Algorithmiq Ltd, Kanavakatu 3C, 00160 Helsinki, Finland}

\author{Ivano Tavernelli} \email{ita@zurich.ibm.com}
\affiliation{IBM Quantum, IBM Research Europe -- Zurich, 8803 R\"{u}schlikon, Switzerland}

\author{Abhinav Kandala}
\affiliation{IBM Quantum, IBM T.J. Watson Research Center, Yorktown Heights, NY 10598, USA}

\author{Sergey N. Filippov}
\affiliation{Algorithmiq Ltd, Kanavakatu 3C, 00160 Helsinki, Finland}

\begin{abstract}
{
Quantum circuits with local unitaries have emerged as a rich playground for the exploration of many-body quantum dynamics of discrete-time systems. 
While the intrinsic locality makes them particularly suited to run on current quantum processors, the task of verification at non-trivial scales is complicated for non-integrable systems. 
Here, we study a special class of maximally chaotic circuits known as dual unitary circuits---exhibiting unitarity in both space and time---that are known to have exact analytical solutions for certain correlation functions. 
With advances in noise learning and the implementation of novel error mitigation methods, we show that a superconducting quantum processor with 91 qubits is able to accurately simulate these correlators. 
We then probe dynamics beyond exact verification, by perturbing the circuits away from the dual unitary point, and compare our results to classical approximations with tensor networks. 
These results cement error-mitigated digital quantum simulation on pre-fault-tolerant quantum processors as a trustworthy platform for the exploration and discovery of novel emergent quantum many-body phases.
}
\end{abstract}

\maketitle

\begin{figure*}
    \centering
    \includegraphics[width=\textwidth]{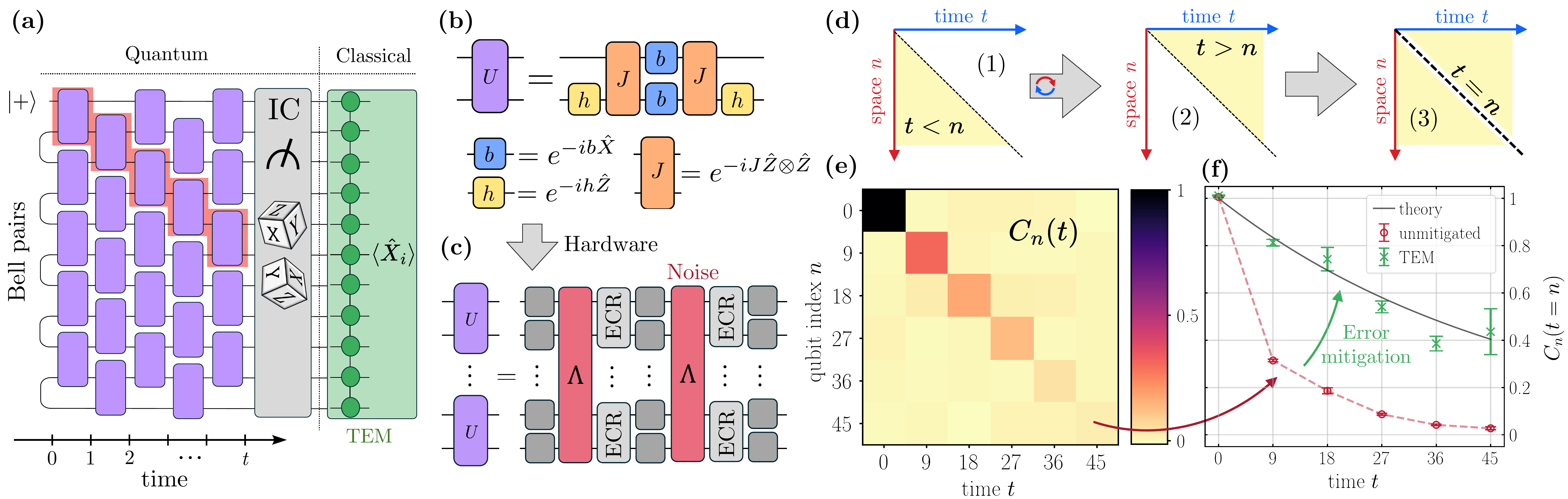}
    \caption[]{
    \textbf{Simulating dual-unitary circuits with tensor-network error mitigation.}
    (a)~Brickwork circuits of dual unitary blocks implement the Floquet evolution of a kicked Ising model. 
    The single-qubit $\hat X_i$ observable on the light cone boundary (red shaded region) yields the infinite-temperature autocorrelation function $C_n(t)$. 
    We take informationally complete (IC) measurements by randomising single-qubit readout between the $\hat X$, $\hat Y$ and $\hat Z$ bases. 
    Measured samples are classically post-processed by the TEM algorithm, which inverts the undesired effects of noise in the circuits.
    (b)~Building blocks of the two-qubit gate $\hat{U}$, which is dual unitary for $\abs{J} = \abs{b} = \pi/4$. 
    (c)~When transpiled to the quantum hardware, one time step consists of two layers of entangling two-qubit gates alternating with single-qubit gates. 
    We use the \emph{echoed cross-resonance gate} (ECR), which is equivalent to the CNOT gate up to local rotations.
    We model noise as Pauli channels $\Lambda$ associated with every unique layer of ECR gates.
    (d)~In brickwork circuits, information spreads in a light cone shape such that all correlations are zero between points $(n_0, t_0) = (0, 0)$ and $(n, t)$ for $t < n$ (1). 
    Similarly, dual unitarity limits information spread in the spatial direction such that correlations vanish  for $n < t$ (2). As a result, non-zero correlations are only found on the boundary of the light cone where $t=n$ (3). 
    (e)~Unmitigated measurements of $C_n(t)$ for a dual unitary circuit exemplified on data for $N=91$ qubits and $h=0.1$.
    (f)~Error mitigation recovers the correct decay of the autocorrelation function.
    }
    \label{fig:Fig1}
\end{figure*}

\begin{figure*}
    \centering
    \includegraphics[width=\textwidth]{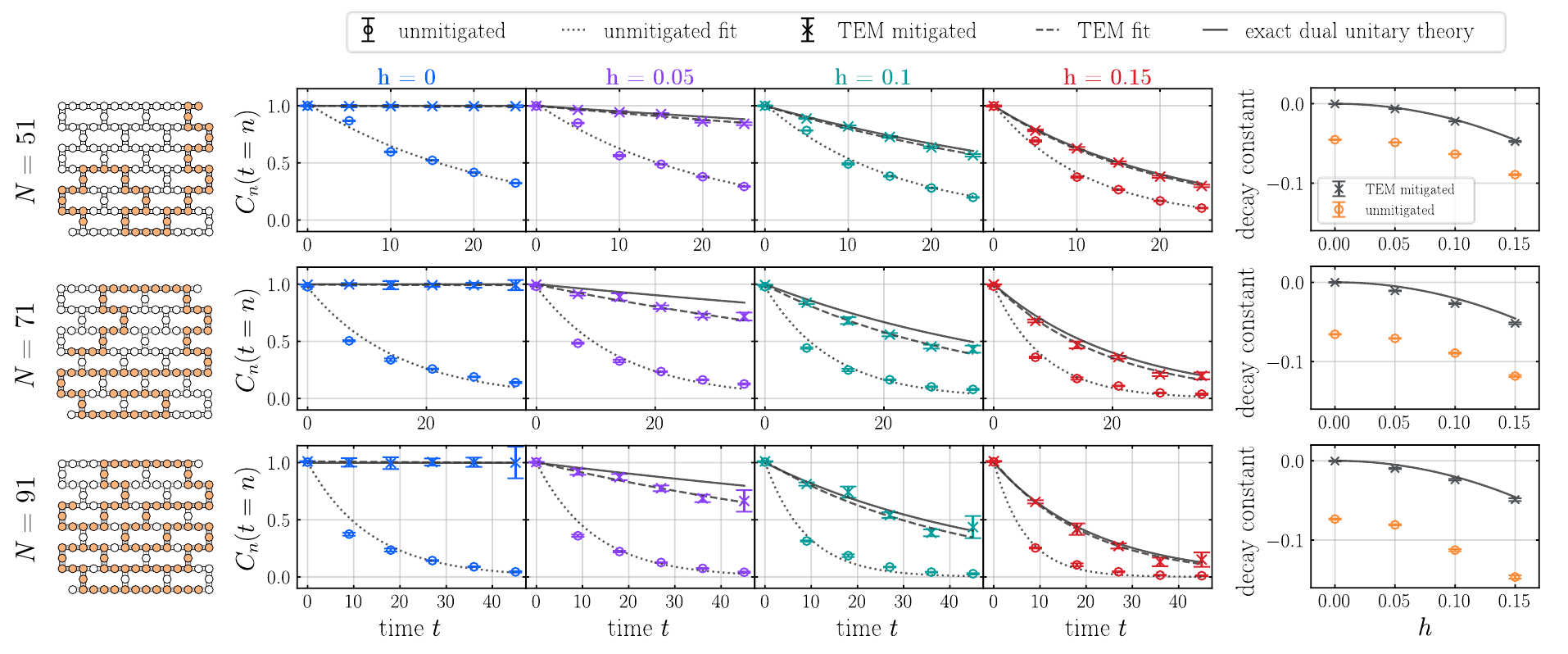}
    \caption[]{
    \textbf{Autocorrelation function at the dual unitary point.} The central four columns depict the experimental autocorrelation function on the light cone $C_n(t=n)$ for increasing values of $h$. 
    The top, middle, and bottom rows correspond to 51-, 71-, and 91-qubit experiments, respectively (qubit layout shown on the left). 
    In each plot, we show the unmitigated (circles) and error-mitigated signals (x-marks), with error bars indicating one standard error, alongside the theoretical curve (black solid lines). 
    For the Clifford point $h=0$, the mitigated signal matches the theoretical curves nearly exactly---as expected, given that the noisy Clifford signal is used to calibrate the noise model (see Methods and Supplementary Information II).
    For $h > 0$, the mitigated points show good agreement with the theoretical curve, albeit with some deviations, particularly for $h=0.05$. 
    We further assess the quality of the results by experimentally inferring the decay rate in each case. 
    We fit exponential curves to the unmitigated and mitigated data (dotted and dashed lines, respectively), and compare the resulting decay rates of the autocorrelation function against the theory in the rightmost column. 
    The results show an excellent agreement between the mitigated values and the theory. 
    }
    \label{fig:Fig2_DU_results}
\end{figure*}

Traditionally, the study of many-body quantum dynamics has been that of continuous time processes. 
In fact, digital quantum simulation algorithms were originally devised as ways of decomposing a continuous evolution into elementary, discrete steps that could be realised on any universal quantum computing architecture~\cite{miessen2023quantum}.
However, as hardware platforms matured and became capable of executing large-scale quantum circuits, a different paradigm emerged. 
In this new scenario, the building blocks of quantum circuits themselves---local unitary gates and measurements---directly give rise to non-equilibrium, discrete-time phenomena. 
Crucially, these protocols can be implemented exactly at any circuit depth, as they are by definition not subject to the algorithmic errors affecting, for instance, the well-known Trotter decompositions of Hamiltonian evolution based on exponential product formulas.
A remarkable example is represented by the simulation of stroboscopic Floquet dynamics, which offers a wealth of new possibilities to probe unexplored universal and emergent phases. 
This includes the investigation of random quantum circuits~\cite{fisher2023random}, computational sampling problems~\cite{hangleiter2023computational}, measurement induced criticality~\cite{fisher18,skinner19}, the emergence of time-crystalline order~\cite{mi2022time}, the existence of many-body localised phases~\cite{mbl1,mbl2,mbl3}, and integrable circuits~\cite{integrable}. 

Among the many advancements that the digitalisation of quantum dynamics brought in the theory of many-body quantum physics and quantum chaos~\cite{chan18}, a key development has been the identification of a class of models known as \emph{dual unitary} (DU) circuits~\cite{Ber-19a}.
These are composed of gates that exhibit unitarity in both the temporal and spatial dimensions.
This unique characteristic allows for the exact computation of certain system properties that would typically be exceedingly challenging to evaluate~\cite{Pir-20a, Ber-20b, Ipp-21a, Suz-22}.
DU circuits act as rapid scramblers of quantum information, with two-time correlation functions and out-of-time correlators propagating at their maximum possible velocities~\cite{Ber-19a, Cla-20a, Ber-20a}, a signature that has already been recorded, e.g., in Ref.~\cite{chertkov2022holographic}.
For this reason, these circuits are often described as ``maximally chaotic''~\cite{Ber-18a, Ber-21a}.
Similarly, for certain solvable initial states, it has been shown that entanglement growth occurs at the maximum rate~\cite{Ber-19b, Zho-22a}.

The ability to simulate Floquet dynamics---including DU circuits---with local gates and short-depth quantum circuits makes them particularly suitable to explore with current, pre-fault-tolerant quantum computers.
Although advances in scale and quality have already enabled the exploration of increasingly complex quantum simulation~\cite{mi2021information,keenan2023evidence,kim2023evidence,robledomoreno2024, shinjo2024unveiling, PhysRevD.109.114510, shtanko2023uncovering} on these processors, their accuracy is still impacted by noise.
Error mitigation~\cite{Temme2017,Li2017,Kandala2019Error,van2023probabilistic, kim2023evidence} has emerged as a powerful tool to extract noise-free observables by post-processing the outputs of several noisy quantum circuits, without the qubit overhead of quantum error correction. 
In this context, error mitigation was recently shown to produce accurate computations from a pre-fault tolerant quantum computer at scales beyond brute-force classical simulation~\cite{kim2023evidence}.
However, a natural question then emerges: in general, how does one build trust in error-mitigated quantum computations at these scales?
While Clifford circuits are a powerful benchmarking tool~\cite{kim2023evidence}, they may not be representative of performance at parameter regimes of interest~\cite{govia2024bounding}.
Dual unitary Floquet models like the one studied in this work can serve as relevant benchmarks in this context, producing non-Clifford circuits of non-trivial scales with analytical solutions.

\vspace{0.5cm}
In this work, we accurately simulate the chaotic dynamics of DU circuits with up to 91 superconducting transmon qubits and 4095 two-qubit gates on \textit{ibm\_strasbourg} and then extend the simulations beyond analytically tractable points. Our results are enabled by our ability to accurately characterise the noise on a large quantum processor, in conjunction with the recently introduced \emph{tensor-network error mitigation} (TEM) method~\cite{filippov2023scalable, filippov2024scalabilityquantumerrormitigation}, that mitigates errors entirely in post-processing, employing tensor networks to implement the inverted noisy channel.
\newpage

{\bf \emph{Dual unitary circuits---}}
Any two-qubit gate is represented by an operator $\hat U$ that satisfies the unitary property $\hat{U}^\dagger \hat{U} = \hat{U} \hat{U}^\dagger = \hat{\mathbb{1}}$. \emph{Dual unitary} (DU) gates are the subset of two-qubit gates with the additional property that they are unitary when viewed as propagators along the spatial direction instead of the temporal direction. 
DU circuits consist of $N$ qubits evolving by a ``brickwork'' pattern of dual-unitary gates, as shown in Fig.~\ref{fig:Fig1}(a), see Methods. 
From the parametrisation of a general two-qubit DU gate, it can be seen that circuits representing the time evolution of certain kicked Ising models are dual-unitary. These circuits are particularly amenable to our hardware, where the native two-qubit interaction is $ZX$, generated by \emph{echoed cross-resonance} (ECR) gates.  
Specifically, we simulate the dynamics of the Ising Hamiltonian $\hat{H}_{\rm I} = J \sum_{n=0}^{N-2}\hat{Z}_{n} \hat{Z}_{n+1} + h \sum_{n=0}^{N-1} \hat{Z}_{n}$, which is periodically ``kicked'' by a transverse field $\hat{H}_{\rm K} = b \sum_{n=0}^{N-1}\hat{X}_{n}$, where $\hat{X}_{n}$, $\hat{Y}_n$, $\hat{Z}_n$ are local Pauli operators on qubit $n$. 
Every time step of the evolution applies the Floquet unitary $\hat{\mathbb{U}}_{\rm KI} = e^{-i\hat{H}_{\rm K}}e^{-i\hat{H}_{\rm I}}$ to the evolved state.
We implement this Floquet evolution through a brickwork circuit, 
where the two-qubit building blocks are specified by the model parameters $h$, $J$, and $b$, as illustrated in Fig.~\ref{fig:Fig1}(b), see Supplementary Information I. 
For $\abs{J} = \abs{b} = \pi/4$, the gates are dual unitary for any choice of $h$~\cite{Aki-16}. 
If $h=0$, the model is integrable, as it can be mapped to free fermions, and the corresponding brickwork circuit is composed of Clifford gates. For a general choice of $h$, the model becomes non-integrable. Yet, analytical solutions exist for the time evolution of certain correlation functions.

Here, we simulate infinite-temperature autocorrelation functions of the form 
\begin{equation}
C_n(t)=\Tr[\hat{\rho}_{\infty} \hat{X}_{0}(0)\hat{X}_{n}(t)],
\label{eq:autocorr}
\end{equation}
where $\hat\rho_{\infty}$ is the infinite-temperature (maximally-mixed) initial state $\hat{\rho}_{\infty}=\mathbb{1}/2^N$, and $t$ denotes the number of time steps. 
Exploiting the dual-unitary property, the autocorrelation function can be calculated exactly~\cite{Ber-19a} for our model as 
\begin{equation} 
C_n (t) = \begin{cases} \left[\cos(2h)\right]^t \, & \text{if } n=t \\ 
            0 \,&\text{otherwise}
		 \end{cases}
\label{eq:C_exact} 
\end{equation} 
for $t \leq (N-1)/2$ (where $N$ is odd), see Supplementary Information I. 
As dual unitarity limits causality to within not only a temporal but also a spatial light cone, the autocorrelation function vanishes outside of the light cone boundary of $n=t$, see Fig.~\ref{fig:Fig1}(d). 
On the light cone, the autocorrelation function is constant at the integrable Clifford point $h=0$, but otherwise decays exponentially in time. 

\begin{figure*}
    \centering
    \includegraphics[width=\textwidth]{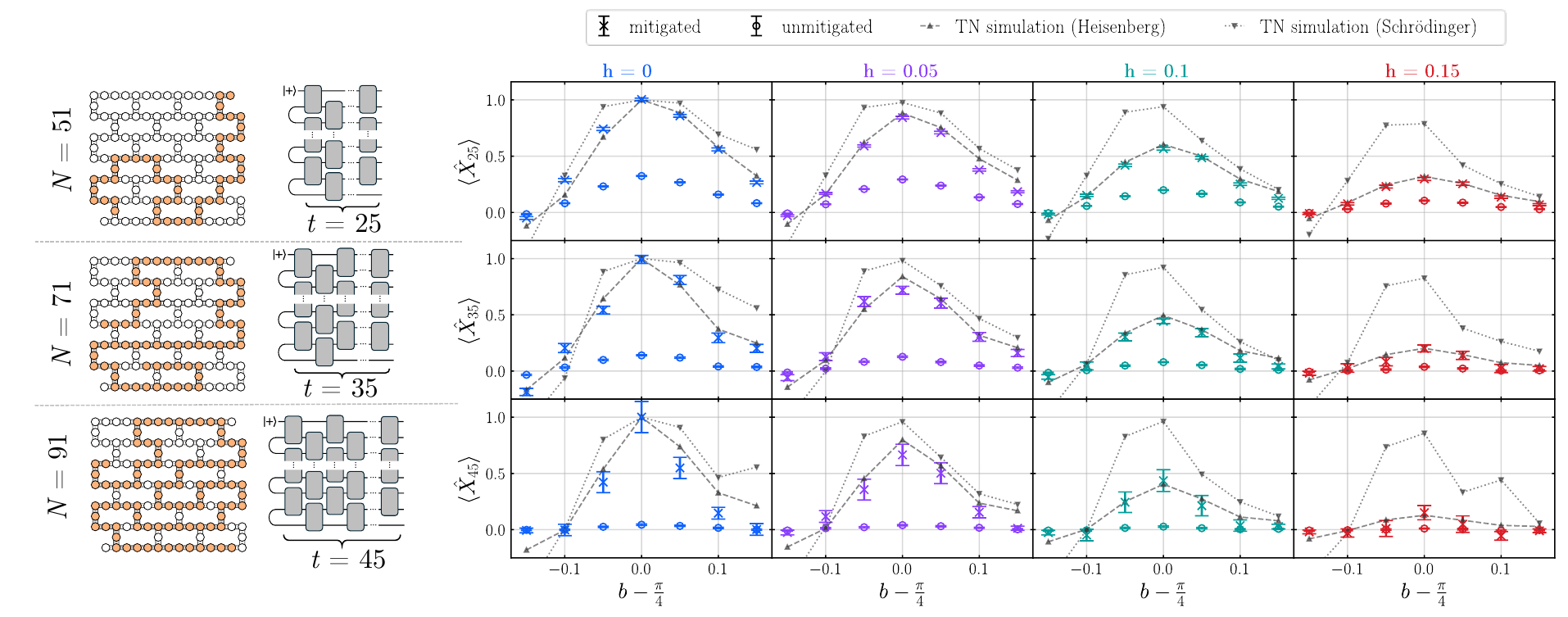}
    \caption[]{\textbf{Non-dual-unitary circuits beyond exact classical verification.} Each plot shows the evolution of $\langle \hat{X}_t (t) \rangle$, with $t = (N - 1) / 2$, as the transverse field $b$ is swept away from dual-unitarity, for a different value of $h$ and system size $N$.
    The dual unitary points $b = \pi / 4$ correspond to the right-most points in Fig.~\ref{fig:Fig2_DU_results}. 
    No analytical solution exists for $b \neq \pi / 4$ and a brute-force statevector simulation is not available either given the scale of the quantum circuits. 
    Instead, we compare our results against classical tensor-network simulations in the Schr\"odinger (dotted line, $\chi=$1500) and the Heisenberg (dashed line, $\chi$=500) pictures. 
    }
    \label{fig:Fig3_beyond_DU}
\end{figure*}

{\bf \emph{Setup---}}
The product of the observables $\hat{X}_0 (0)$ and $\hat{X}_n(t)$ taken at two different points in time in Eq.~\eqref{eq:autocorr} makes experimental access to $C_n(t)$ a non-trivial task.
However, at the dual unitary point, the infinite-temperature autocorrelation function can be rewritten as an expectation value $C_n(t) = \bra{\Psi (0)} \hat{X}_n (t) \ket{\Psi (0)}$, where the initial pure state $\ket{\Psi(0)} = \ket{+}_0 \otimes \ket{\psi_{\rm Bell}}^{\otimes \lfloor(N-1)/2 \rfloor}$ prepares the $0$-th qubit in $\ket{+}_0=(\ket{0}_0+\ket{1}_0)/\sqrt{2}$ and all other qubits in a product of Bell pairs $\ket{\psi_{\rm Bell}} = (\ket{00}+\ket{11})/\sqrt{2}$ (for all qubit pairs $i, i+1$, with $i=1,3, \dotsc, N-2$), as illustrated in Fig.~\ref{fig:Fig1}(a), see Supplementary Information I. Therefore, at the dual-unitary point, the task of estimating $C_n(t)$ conveniently reduces to measuring $\expect{\hat X_n}$ at the end of the circuit of Fig.~\ref{fig:Fig1}(a).

The measurement of $C_n(t)$ for various qubits $n$ and time steps $t$ is shown in Fig.~\ref{fig:Fig1}(e). 
As predicted analytically, we observe a negligible signal for $t\neq n$ and finite signal is only measured along the light-cone boundary for $t=n$. 
However, as a consequence of noise, the measured autocorrelation function on the light cone boundary decays quicker with $t$ than the exact evolution from Eq.~\eqref{eq:C_exact}. 
To mitigate these detrimental effects of noise, we rely on the recently developed TEM method~\cite{filippov2023scalable}, see Fig.~\ref{fig:Fig1}(a).

The basic idea behind TEM is to construct an approximate, efficient tensor network representation $\mathcal{M'}$ of the noise cancelling map $\mathcal{M}$, which maps the noisy state produced by the device to the ideal noiseless state, $\mathcal{M}(\hat \rho_{\rm noisy}) = \hat \rho_{\rm ideal}$. 
The map is then used to estimate the noiseless expectation value of an observable $\hat O$ as ${\rm Tr}[\hat \rho_{\rm ideal} \hat O] \approx {\rm Tr}[\hat \rho_{\rm noisy} \mathcal{M}^{\prime\dagger}(\hat O)]$, that is, by measuring the expectation value of the observable $\hat O' = \mathcal{M}^{\prime\dagger}(\hat O)$ on the noisy state $\hat \rho_{\rm noisy}$.
While $\hat O'$ can generally have a non-trivial support over a vast number of Pauli strings, it is possible to obtain unbiased estimators of its expectation value by using informationally complete measurements, realised through randomised selection of readout bases. 
We sample the measurement bases uniformly at random for each qubit, except for the signal qubit $i=t$, where the observable is biased towards $\hat X$ ($80\%$ probability for $\hat X$, $10\%$ for $\hat Y$, and $10\%$ for $\hat Z$), as the estimated observable is dominated by the $\hat X$-contribution, see Methods and Supplementary Information III.

The approximate noise-cancelling map $\mathcal{M}$ relies on a representative model of the device noise.
We tailor the noise of each layer of ECR gates (see Fig.~\ref{fig:Fig1}(c)) with Pauli twirling~\cite{twirling_bennet, twirling_knill, wallman2016noise,Hashim2021} and characterise the resulting Pauli channels by building on the noise learning technique established in Refs.~\cite{kim2023evidence, van2023probabilistic}.
As a novel extension of this technique, we use the Clifford point of the DU circuits ($h=0$) to fine-tune previously unconstrained degrees of freedom of the noise model~\cite{van2023probabilistic,chen2023learnability, chen2024efficientselfconsistentlearninggate} (see Methods and Supplementary Information II). With this machinery in place, the TEM-mitigated values $C_n(t)$ retrieve the predicted decay of the autocorrelation function, see Fig.~\ref{fig:Fig1}(f).  

{\bf \emph{Results---}} 
Using the above approach, we first simulate the infinite-temperature autocorrelation function at various dual unitary points. 
We consider several values of the field $h \in \{0, 0.05, 0.1, 0.15\}$ and benchmark the performance at different system sizes of 51, 71, and 91 qubits, see Fig.~\ref{fig:Fig2_DU_results}. 
Even when integrability is broken for $h > 0$, we are able to closely recover the expected behaviour for all considered values of $h$. 
At larger system sizes, we report small deviations in the mitigated results.
As the system size increases, note that the circuit depth also increases, and at these larger circuit volumes, errors in the noise model can accumulate, leading to a residual bias in the mitigated results~\cite{govia2024bounding}. 
These are likely a consequence of, for instance, imperfections in noise learning, model violations due to incorrect model assumptions, or even the increased instability in the noise from the longer runtimes associated with larger circuit volumes~\cite{kim2024error}. Benchmarking the accuracy of the measured noise model in predicting the experimental noisy data for other families of Clifford circuits reveals small systematic errors, that are particularly prominent at longer depths (see Supplementary Information II E and II F).

The decay rate of $C_{n}(t)$ as a function of $h$ given in Eq.~\eqref{eq:C_exact} is a universal quantity independent of the system size. We demonstrate that with error mitigation we can accurately recover the exact prediction of this decay constant for all system sizes studied. 
This not only showcases the effectiveness of our approach in studying high-temperature autocorrelation functions of large-scale quantum chaotic circuits but also provides a valuable benchmark of system performance for non-Clifford circuits.

After assessing the accuracy of the mitigated results for analytically solvable DU circuits, we perturb away from the DU parameters. 
We note that, while working with the same initial state and observable as before, the local expectations values $\expect{\hat{X}_n(t)}$ lose their interpretation as autocorrelation functions away from the DU point. 
In Fig.~\ref{fig:Fig3_beyond_DU}, at each of the previously considered values of $h$,
we report the change of $\expect{\hat{X}_n(t)}$ for the final simulation time $n = t =(N-1)/2$ as we perturb the transverse field $b$ away from dual unitarity. 
We reiterate that in the absence of exact analytical solutions and at a scale beyond brute-force classical simulation, these computations can only be compared to approximate classical methods. 

Here, we compare the error-mitigated results to tensor network simulations in both the Heisenberg  and Schr\"odinger pictures.  
Across the different parameters, the experimental data show strong agreement with the Heisenberg simulations with some deviations arising at larger circuit volumes, but large disagreements with the Schrödinger-picture simulations. 
The Heisenberg-picture simulations in Fig.~\ref{fig:Fig3_beyond_DU} employ bond dimension $\chi=500$ and display evidence for convergence at smaller bond dimension as well (see Supplementary Information IV~B and IV~C). 
In contrast, the Schr\"odinger-picture simulations are seen to have not converged even at bond dimension $\chi=1500$ (see Supplementary Information IV~A and IV~C). Therefore, while dynamics in the Heisenberg picture are converging on classical computers, simulations in the Schr\"odinger picture become unaffordable at the scale of our experiments. In our quantum-classical workflow, to produce the error-mitigated data points for $N = 91$, the middle-out contraction for TEM employs bond dimension $\chi=70$. We emphasize that even in the limit of infinite bond dimension, the accuracy of TEM can generally be limited to the accuracy of the quantum component.

Furthermore, for the quantum component of the workflow, the traces of data shown in Figs.~\ref{fig:Fig2_DU_results} and~\ref{fig:Fig3_beyond_DU} for $N=91$, including noise learning and mitigation, took a wall clock time of \qty{3}{\hour}~\qty{24}{\minute}.
This involved taking 262,144 individual shots per data point, at a sampling rate exceeding \qty{1}{\kilo\hertz}, enabled by fast parametric circuit compilation to perform gate twirling and readout basis randomisation with 256 circuit instances, see Supplementary Information II~B.
This marks a significant improvement over the $\mathcal{O}(\qty{10}{\hertz})$ rates reported in previous experiments~\cite{kim2023evidence}.
Our results emphasise the progress of error-mitigated quantum computing in becoming increasingly competitive with widely used classical algorithms in regimes where brute-force exact solutions are unavailable.
 
{\bf \emph{Discussion---}} 
The framework, methodology and results displayed in this work highlight the utility of pre-fault-tolerant quantum processors for studying models at the forefront of quantum many-body physics. 
First and foremost, we demonstrate the capability to accurately simulate the decay of autocorrelators at the dual unitary point of the kicked Ising model. 
We believe that our work will inspire further experiments of condensed matter physics where the same autocorrelation functions can be used to extract transport properties~\cite{prosen_bal} and predict the existence of localised phases~\cite{mbl_phenom}.
Secondly, by leveraging the analytic tractability of dual unitary circuits, we demonstrate how these systems can serve as performance benchmarks for non-Clifford circuits. 
Thirdly, and perhaps most importantly, we advance the boundaries of quantum simulation on multiple technical fronts. 
Central to our approach is the integration of quantum and classical resources, achieved through the implementation of TEM~\cite{filippov2023scalable}. 
To this end, we run an accurate characterisation of the device noise channels which improves on previously established noise learning techniques~\cite{van2023probabilistic, kim2023evidence}.
Our experiment adds to the growing body of work that leverages classical computation to extend the reach of near-term quantum processors~\cite{filippov2023scalable,robledomoreno2024,eddins2024,robertson2024}. 
As quantum hardware advances towards lower error rates~\cite{stehlik2021tunable}, more stable noise~\cite{kim2024error} and faster speeds~\cite{wack2021,rajagopala2024}, our approach could open up the path to the first class of quantum simulations of many-body dynamics on universal quantum processors that surpass classical simulators already before the advent of fault-tolerance.

\noindent\rule{\linewidth}{1pt}

\section*{Methods}

{\bf {Dual unitarity}}
Given a two-qubit unitary $\hat{U} = \sum_{i,j,k,l=0}^1 U_{ij}^{kl} \ket{k}\bra{i}\otimes\ket{l}\bra{j}$, one defines a dual operator $\hat{U}_D = \sum_{ijkl} U_{ij}^{kl} \ket{j}\bra{i}\otimes\ket{l}\bra{k}$ through a shuffling of some input/output subsystems of $\hat{U}$ (exchange of the bra/ket indices $j \leftrightarrow k$). 
If the dual is unitary, i.e. if $\hat{U}_D^{\dagger} \hat{U}_D = \hat{U}_D \hat{U}_D^{\dagger} = \hat{\mathbb{1}}$, then the gate $\hat{U}$ is called \emph{dual-unitary}. 
Dual-unitary circuits, which will be the primary focus of our simulations, consist of $N$ qubits (labelled $n = 0,1,\hdots,N-1$) evolving by a ``brickwork'' pattern of dual-unitary gates $\hat{U}_{n,n+1}$. The brickwork is an even layer of dual-unitary gates $\hat{\mathbb{U}}_e = \bigotimes_{j=0}^{(N-1)/2-1} \hat{U}_{2j,2j+1}$ followed by an odd layer $\hat{\mathbb{U}}_o = \bigotimes_{j=1}^{(N-1)/2} \hat{U}_{2j-1,2j}$, repeated periodically. We define our unit of time to be the evolution by single layer, odd or even, so that the brickwork Floquet unitary $\hat{\mathbb{U}} = \hat{\mathbb{U}}_o \hat{\mathbb{U}}_e$ evolves the system through two units of time.
\medskip

{\bf {Noise characterisation}}
We model noise as a sparse Pauli-Lindblad channel associated with twirled Clifford layers of parallel ECR gates, building on Refs.~\cite{van2023probabilistic, kim2023evidence}.
We choose the convention of the noise channel $\Lambda$ acting before the $n$-qubit unitary layer $\hat U$.
The Lindblad generator $\mathcal{L}$ of this noise channel $\Lambda=\mathrm{e}^{\mathcal{L}}$ has the form
\begin{equation}
\label{eq:noise_model_def}
\mathcal{L}(\hat \rho) =   \sum_{i} \lambda_i \Bigl( \hat P_i\hat  \rho \hat P_i^\dagger - \hat \rho \Bigr),
\end{equation}
where $\lambda_i$ are the generator rates associated with Pauli jump operators $\hat P_i$ and $i$ indexes the set of all single-qubit and nearest-neighbour two-qubit Pauli strings to capture crosstalk of neighbouring ECR gates. 
We calibrate this noise model by fitting the generator rates to measured Pauli fidelities $f_i = \Tr\bigl(\hat P_i {\Lambda}(\hat P_i) \bigr) / 2^n$ following Ref.~\cite{van2023probabilistic}.
However, due to a fundamental gauge degree of freedom, this protocol does not distinguish between the fidelity of a given Pauli $ \hat P_a$ and its conjugate $\hat P_{a^\prime} = \hat U\hat P_{a}\hat U^\dagger$~\cite{chen2023learnability}.
Instead, we obtain the pair fidelities $\overline{f_a} := \sqrt{f_{a} f_{a^\prime}}$.
In previous work, the generator rates $\lambda_i$ were fit directly to the pair fidelities implicitly assuming $f_a = f_{a^\prime}$. 

In this work, we move beyond this assumption with the key idea of treating the Clifford point of the kicked Ising evolution ($h=0$) as additional learning circuits, see Supplementary Information II~D.
Up to SPAM errors, the noisy signal is a product of Pauli fidelities, i.e., $\expect{ \hat X_j }_{\text{noisy}} = f_1^\text{C} \dotsm f_{2j}^\text{C}$.
The contributing fidelities $f^C_i$ form a subset of the sparse Pauli basis, alternating between single-qubit and two-qubit Paulis as the signal travels along the boundary of the light cone. 
We introduce weights $\alpha_i$ such that $f^C_i(\alpha_i) = \alpha_i \overline{f^C_i}$ and $f^C_{i^\prime}(\alpha_i) = \overline{f^C_i} / \alpha_i$.
The values of $\alpha_i$ are chosen such that the contributing fidelities match the observed signal $\langle \hat X_j \rangle_\text{noisy}$, after applying twirled readout error mitigation~\cite{van2022model}.
The vectorised noise generators $\boldsymbol{\lambda}$ are finally obtained through a least-square fit to the vectorised fidelities $\boldsymbol{f}$ and conjugate fidelities $\boldsymbol{f^\prime}$ by solving 
\begin{equation}
\label{eq:generator_fidelities_fit}
\operatorname*{arg\,min}_{\lambda_i \geq 0} \; \Bigg\lVert
\left[\begin{array}{c}M \\ M^\prime \end{array}\right] \boldsymbol{\lambda}
+ \frac{1}{2} \log \left[\begin{array}{c} \boldsymbol{f}(\boldsymbol{\alpha}) \\ \boldsymbol{f^\prime}(\boldsymbol{\alpha}) \end{array}\right]  \Bigg\rVert_2^2
\end{equation}
where $M$ is the anticommutation matrix of the sparse Pauli basis with $M_{ij} = 1$ if $ \hat P_i$ and $\hat P_j$ anticommute (otherwise $M_{ij} = 0$), and similarly $ M^\prime$ for the conjugate Pauli basis. 
For those fidelities not captured by the Clifford kicked Ising circuits, we keep the assumption of symmetric fidelities (i.e., $\alpha_i = 1$). 
We note that all Pauli fidelities remain $\leq 1$, as required for a physical noise channel.
By construction, the resulting noise model is in perfect agreement with the experiments performed at the Clifford point, resulting in the perfect mitigation in the leftmost column of Fig.~\ref{fig:Fig2_DU_results}.
The non-Clifford DU circuits serve as an independent classically-verifiable benchmark. 
\medskip

{\bf {Tensor-network error mitigation}} The \emph{tensor-network error mitigation} (TEM) algorithm works by inverting the noise inherent in the quantum device during the classical post-processing stage, undoing its effects without altering the dynamics on the quantum hardware~\cite{filippov2023scalable}. 
This method has been numerically shown to be highly effective in providing mitigated estimates of observables and achieves the universal lower bound for sampling overhead in noise mitigation methods for stochastic noise under relevant experimental conditions~\cite{filippov2024scalabilityquantumerrormitigation, Tsubouchi2023costbound}.

The structure of the quantum circuit, depicted in Fig.~\ref{fig:Fig1}(a) consists of a series of ideal unitary layers ${\cal U}_l = \hat{U}_l \bullet \hat{U}_l^{\dag}$, each preceded by an associated noise layer \( \Lambda_l \) of the sparse Pauli-Lindblad form. The map
\begin{equation}
\label{eq:TEM_def}
{\cal M} = (\bigcirc_l {\cal U}_l) \circ \bigcirc_l (\Lambda_l^{-1} \circ {\cal U}_l^{-1} )
\end{equation}
undoes the effect of noise, when applied to the output density operator $\hat \rho$ of a noisy computation, i.e., ${\rm Tr}[{\cal M}(\hat\rho)\hat O] = {\rm Tr}[\hat\rho {\cal M}^{\dag}(\hat O)] = \expect{\hat O}_{\rm ideal}$ for any observable $\hat O$. The operator ${\cal M}^{\dag}(\hat O)$ is the TEM-modified observable, whose average value on the noisy state $\hat \rho$ gives the noise-free estimation of the original observable $\hat O$. 

Constructing the map $\cal M$ as in Eq.~\eqref{eq:TEM_def} by concatenating the constituent maps layer by layer would lead to a complexity growing exponentially in the number of layers. However, the computation is made efficient via the recurrence relation
\begin{equation}
    {\cal M}_l = {\cal U}_l \circ \Lambda_{l}^{-1} \circ {\cal M}_{l-1} \circ {\cal U}_{l}^{-1},
    \label{Tem_iteration}
\end{equation}
where every unitary layer ${\cal U}_l$ and corresponding inverse noisy layer $(\Lambda_{l}^{-1} \circ {\cal U}_l^{-1})$ approximately cancel each other. Both ${\cal U}_l$ and $\Lambda_l^{-1}$ allow a tensor network representation in the form of the \emph{matrix product operator} (MPO) of bond dimension $4$~\cite{filippov2023scalable}, so the map ${\cal M}$ is constructed via recurrent conventional tensor network contractions of MPOs, with ${\cal M}_0 = {\rm Id}$ being the identity transformation. Compression of the MPO at each iteration results in the approximate map ${\cal M}'$ capturing the most significant contributions (Supplementary Information III~B).

One of the ways to measure the TEM-modified observable ${\cal M}'^{\dag}(\hat O)$ is to make use of \emph{informationally complete} (IC) measurements at the end of the quantum processing unit (Supplementary Information III~C). In this study, IC measurements are implemented through qubit-wise randomised projective measurements in the eigenbasis of either of the Pauli operators. Measurements are accompanied by twirled readout error mitigation based on random Pauli bit flips before the standard measurement, which enables us to account for readout noise through a single multiplicative factor for each Pauli operator~\cite{van2022model}. Given the TEM-modified observable ${\cal M}'^{\dag}(\hat O)$ and the outcomes of IC measurements for a noisy output $\hat \rho$ of a quantum processing unit, the estimation of the mitigated signal ${\rm Tr}[\hat \rho {\cal M}'^{\dag}(\hat O)]$ is obtained via tensor network machinery (Supplementary Information III~D). The estimation accuracy is observable dependent, so in practice one can exploit additional degrees of freedom and symmetries in the circuit to reduce the measurement cost for the resulting TEM-modified observable (Supplementary Information III~E). Stochastic errors in the noise-mitigated signal are compared between TEM and other noise mitigation techniques in Supplementary Information III~F. 

\medskip
\textbf{Classical simulation} \label{sec:simulations}
We benchmark the noise-mitigated results against purely classical approximate simulations of the noiseless circuits in both the Schrödinger and Heisenberg picture by using tensor network techniques (Supplementary Information~IV).
Simulations in the Schrödinger picture are particularly demanding and inefficient, because the initially prepared local correlations in the form of Bell pairs quickly become highly non-local due to the entangling nature of the circuit (Supplementary Information IV~A).
These simulations remain far from convergence even with the high bond dimensions employed (\num{1500}).
In contrast, simulations in the Heisenberg picture provide reliable estimations with more moderate resources: the absolute difference between results with bond dimension \(\chi\) and \(\chi+100\) is below \(10^{-2}\) for \(100\leq\chi\leq500\) and below \(10^{-3}\) for \(500\leq\chi\leq900\).
Therefore \(\chi=100\) already provides a satisfactory estimate 
(Supplementary Information IV~B and IV~C).
This level of accuracy can be attributed to the experiments being a perturbation of the Clifford point (\(h=0\), \(b=\tfrac{\pi}{4}\)), at which \(\chi=1\) suffices for the exact simulation in the Heisenberg picture.
We employ several convergence checks to ensure a sufficiently high bond dimension is used beyond the Clifford point to achieve the desired accuracy (Supplementary Information IV~C).
An overview of the required computational resources is given in Supplementary Information III~H.

We also simulate the noisy circuit dynamics with the learned noise models, providing the noisy signal we would expect to measure as an unmitigated result.
This allows us to assess the performance of noise characterisation, and how it affects error mitigation (Supplementary Information III~F).  

\noindent\rule{\linewidth}{1pt}
{\bf Acknowledgements---}
M.R., Z.Z., G.G.P, J.G.~thank Lorenzo Piroli for useful discussions. A.E. , Y.K. and A.K. thank Sergey Bravyi for introducing dual unitary circuits to them. 
We thank Rajeev Malik for enabling device access. 
We thank Alireza Seif, David Layden, Ewout van den Berg, and Luke Govia for feedback on the manuscript. 
I.T., F.T. and L.E.F. thank Stefan Wörner, Almudena Carrera Vázquez, and Daniel Egger for helpful discussions. 
We acknowledge EuroHPC Joint Undertaking for awarding us access to Leonardo at CINECA, Italy and Karolina at IT4Innovations, Czech Republic.
L.E.F. acknowledges funding from the European Union’s Horizon 2020 research and innovation program under the Marie Sk\l{}odowska-Curie grant agreement No.~955479 (MOQS – Molecular Quantum Simulations). 
This research was supported by the NCCR MARVEL, funded by the Swiss National Science Foundation.
S.D. acknowledges support through the SFI-IRC Pathway Grant 22/PATH-S/10812. 
J.G. is supported by a Royal Society University Research Fellowship and would like to thank Silvia Pappalardi for inspirational discussions. 

\bigskip
{\bf Competing interests---}
Elements of this work are included in patent applications filed by Algorithmiq Oy with the European Patent Office and the US Patent Office.

\let\oldaddcontentsline\addcontentsline
\renewcommand{\addcontentsline}[3]{}
\let\addcontentsline\oldaddcontentsline

\clearpage
\newpage
\widetext

\begin{center}
\textbf{\large Supplementary Information:\\ \vspace{0.5cm} Dynamical simulations of many-body quantum chaos on a quantum computer}
\end{center}

\tableofcontents

\setcounter{equation}{0}
\setcounter{figure}{0}
\setcounter{table}{0}
\setcounter{page}{1}
\setcounter{section}{0}
\makeatletter
\renewcommand{\theequation}{S\arabic{equation}}
\renewcommand{\thefigure}{S\arabic{figure}}
\renewcommand{\thesection}{S\Roman{section}}

\clearpage

\section{Theory}
\label{sec:theory}
\subsection{Mapping the kicked Ising model to a brickwork circuit}
\noindent The kicked Ising model is described by the time-dependent Hamiltonian 
\begin{equation} 
\label{eq:kickising} 
\hat{H}_{KI}(t) = \hat{H}_{I} + \sum_{m \in \mathbb{Z}} \delta(t-m)\hat{H}_{K} , 
\end{equation} 
where $\hat{H}_{I} = J \sum_{n=0}^{N-1}\hat{Z}_{n} \hat{Z}_{n+1} + h \sum_{n=0}^{N-1} \hat{Z}_{n}$ is the Ising Hamiltonian, and the system is periodically kicked (with unit period) by the transverse field Hamiltonian $\hat{H}_{K} = b \sum_{n=0}^{N-1}\hat{X}_{n}$.
The Floquet unitary for the stroboscopic evolution generated by $\hat{H}_{KI}(t)$ is 
\begin{equation} 
\label{eq:Ising_Floquet_operator} 
\hat{\mathbb{U}}_{KI} = \mathcal{T}\exp \Big[ -i\int^{1}_{0}\hat{H}_{KI}(t) dt \Big] = e^{-i\hat{H}_{K}}e^{-i\hat{H}_{I}} , 
\end{equation} 
where $\mathcal{T}$ is the time-ordering operator.
As a quantum circuit, this can be written as  
\begin{equation} 
\hat{\mathbb{U}}_{KI} = \exp\left[-ib\sum_{n}\hat{X}_n \right] \exp\left[-J\sum_n \hat{Z}_n \hat{Z}_{n+1} \right] \exp\left[-ih\sum_n \hat{Z}_n \right] = \adjincludegraphics[valign=c]{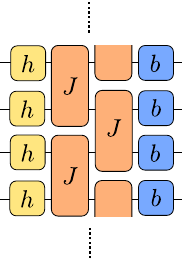} , 
\label{eq:U_KI} 
\end{equation} where we have introduced the gates \begin{equation}  
\adjincludegraphics[valign=c]{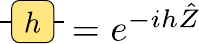} \qquad \qquad \adjincludegraphics[valign=c]{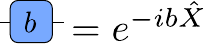} \qquad \qquad \adjincludegraphics[valign=c]{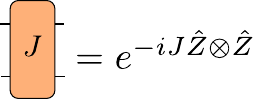}. 
\end{equation} 
Note that, in terms of Pauli rotation gates, $e^{-ih\hat Z} = R_Z(2h)$, $e^{-ib\hat X} = R_X(2b)$, and $e^{-iJ\hat Z \otimes \hat Z} = R_{ZZ}(2J)$.
In this Section we show how the Floquet unitary $\hat{\mathbb{U}}_{KI}$ can be rewritten as a sequence of odd and even layers of a Floquet brickwork circuit.
First, we consider the two-qubit gate 
\begin{equation} 
\hat{U}_{n,n+1} = e^{-ih\hat{Z}_{n}} e^{-iJ\hat{Z}_{n} \hat{Z}_{n+1}} e^{-ib( \hat{X}_{n} + \hat{X}_{n+1}) } e^{-iJ\hat{Z}_{n} \hat{Z}_{n+1}} e^{-ih\hat{Z}_{n}} = \adjincludegraphics[valign=c]{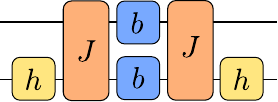} . \label{eq:two_qubit_gate} 
\end{equation} 
A brickwork circuit made up of this gate consists of an even layer $\hat{\mathbb{U}}_e = \prod_{n {\rm ~even}}\hat{U}_{n,n+1}$ and an odd layer $\hat{\mathbb{U}}_o = \prod_{n {\rm ~odd}}\hat{U}_{n,n+1}$ of gates, repeated periodically. A single time step of the brickwork has the circuit 
\begin{equation} 
\label{eq:brickwork_timestep}
\hat{\mathbb{U}} = \hat{\mathbb{U}}_o \hat{\mathbb{U}}_e = \quad \adjincludegraphics[valign=c]{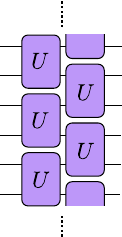} \quad = \quad \adjincludegraphics[valign=c]{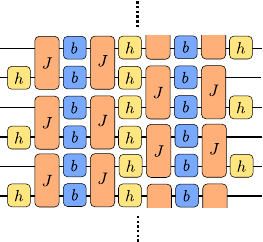} . 
\end{equation}

Define the unitary operator 
\begin{equation} 
\hat{\Sigma} = \prod_{n {\rm ~odd}}e^{-iJ\hat{Z}_n \hat{Z}_{n+1}} \prod_{n {~odd}}e^{-ih\hat{Z}_n} = \quad \adjincludegraphics[valign=c]{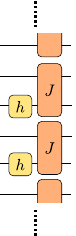} , 
\end{equation} 
and consider the circuit given by 
\begin{equation} 
\hat{\Sigma}^\dagger \hat{\mathbb{U}} \hat{\Sigma} = \quad \adjincludegraphics[valign=c]{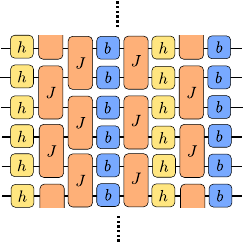} . 
\end{equation} 
Using the fact that the $h$-gates and the $J$-gates commute, and comparing with the circuit diagram for $\hat{\mathbb{U}}_{KI}$ in Eq.~\eqref{eq:U_KI}, we can see that $ \hat{\Sigma}^\dagger \hat{\mathbb{U}} \hat{\Sigma}$ is equal to two periods of the Floquet unitary of the kicked Ising model, i.e., 
\begin{equation}  
\hat{\Sigma}^\dagger \hat{\mathbb{U}} \hat{\Sigma} = \hat{\mathbb{U}}_{KI} \hat{\mathbb{U}}_{KI} . 
\label{eq:relating_U_U_KI} 
\end{equation} 
This shows that the Floquet unitary for the kicked Ising model can be related to a brickwork circuit with a fixed two-qubit gate given by Eq.~\eqref{eq:two_qubit_gate}. 
Since we define our unit of time as the evolution by a single layer of the brickwork circuit (odd or even), Eq.~\eqref{eq:relating_U_U_KI} confirms that a period of the brickwork circuit takes two time steps, while a period of the kicked Ising model takes a single time step (half the period of the brickwork circuit).

\subsection{Infinite temperature correlator }

\begin{figure*}[h!]
    \centering
    \includegraphics[width=0.7\columnwidth]{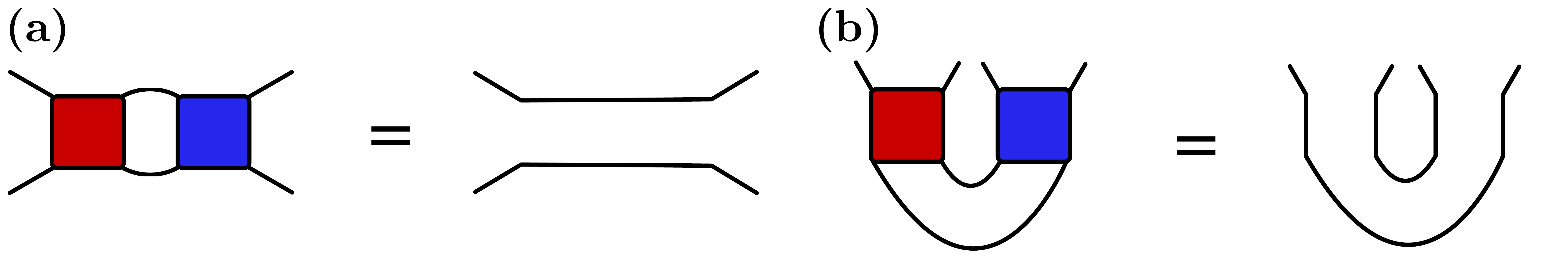}    
    \caption[]{\textbf{Contraction rules for dual unitary gates.} Let the red tensor be a two-qubit dual unitary operator $\hat{U} = \sum_{i,j,k,l=0}^1 U_{ij}^{kl} \ket{k}\bra{i}\otimes\ket{l}\bra{j}$ and the blue tensor its hermitian conjugate $\hat{U}^\dagger$. (a) Diagrammatic representation of the unitary relation $\hat{U} \hat{U}^{\dagger} = \hat{\mathbb{1}}$. (b) Diagrammatic representation of the dual unitary relation $\hat{U}_D \hat{U}_D^{\dagger} = \hat{\mathbb{1}}$ where $\hat{U}_D = \sum_{ijkl} U_{ij}^{kl} \ket{j}\bra{i}\otimes\ket{l}\bra{k}$.
    See Ref.~\cite{Ber-19a} for an introduction to the diagrammatic representation. 
    }
    \label{fig:unitary}
\end{figure*}

Here, we show how the expectation value $\langle \Psi(0)|\hat{X}_n(t)|\Psi(0)\rangle$ where $\ket{\Psi(0)} = \ket{+_0} \otimes \ket{\psi_{\rm Bell}}^{\otimes (N-1)/2 }$ (with $N$ being odd) allows us to calculate the infinite-temperature correlator $C_n(t) = \Tr\bigl(\hat{X}_0\hat{X}_n(t) \bigr)/2^N$ through a diagrammatic representation. 
We start with the case where $n=t$, from the expression 
\begin{equation}
\centering
    2^{\frac{N+1}{2}}\langle \Psi(0)|\hat{X}_n(t)|\Psi(0)\rangle = \adjincludegraphics[valign=c, width=0.2\columnwidth]{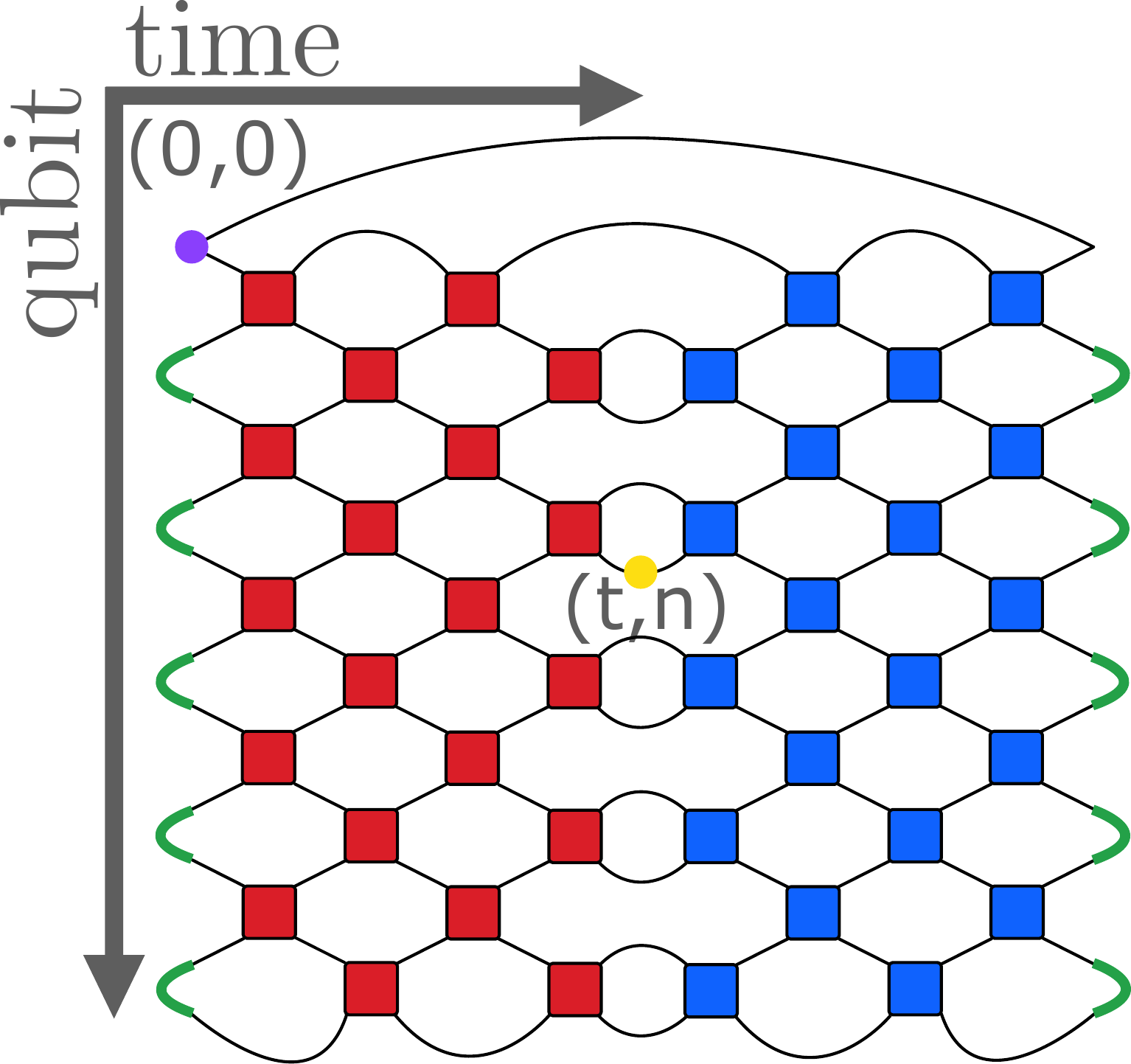} = \quad \adjincludegraphics[valign=c, width=0.2\columnwidth]{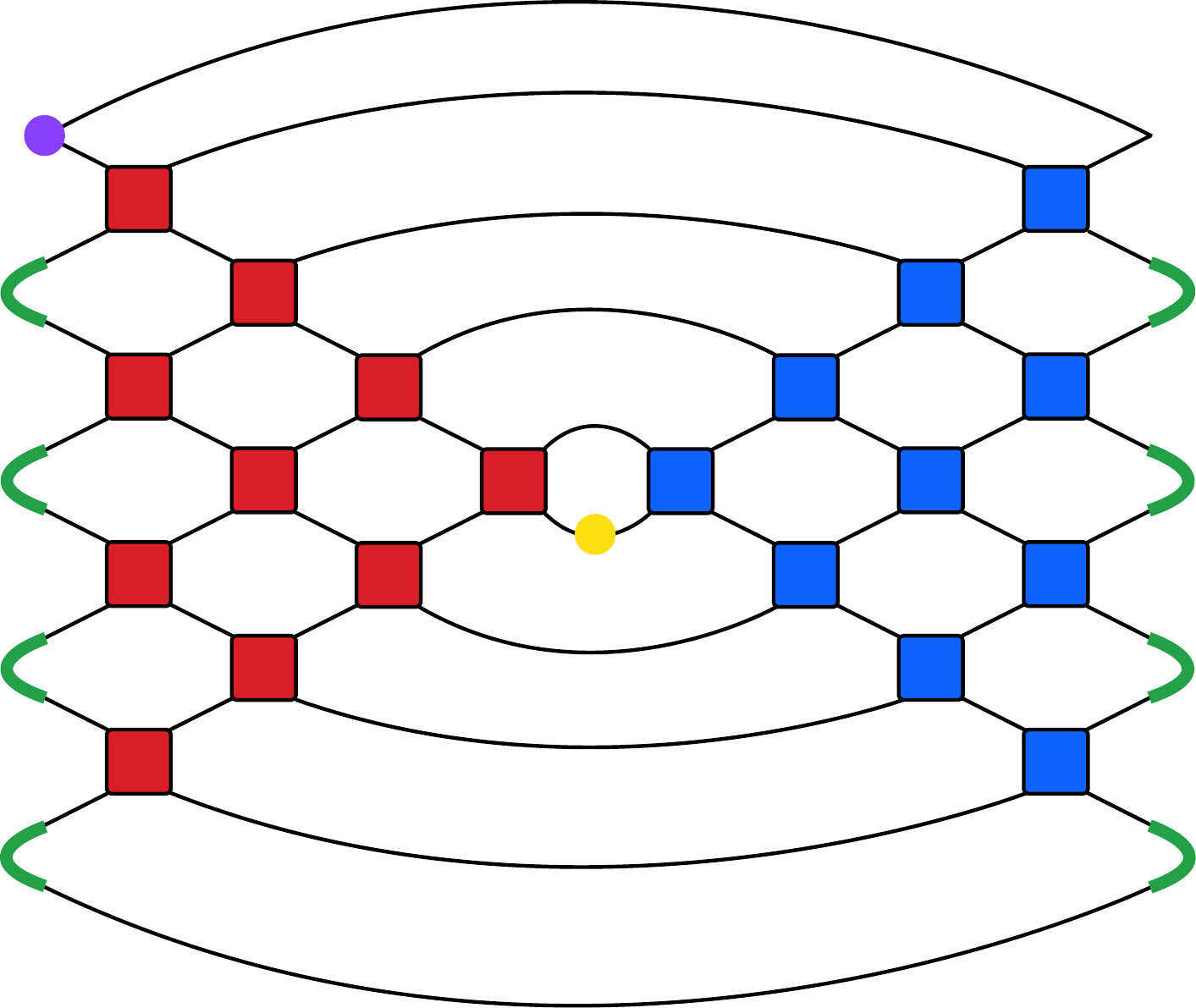}\label{eq:Bell_expect}
\end{equation}
where the yellow tensor at coordinate $(t, n)$ denotes the $\hat{X}$ observable, and the purple tensor at coordinate $(0, 0)$ denotes the state $\ket{+}\bra{+}$. 
The Bell pairs in $\ket{\Psi(0)}$ are shown capping off the left and right sides of the diagram in green. 
We employ the unitary and dual unitary relations shown in Fig.~\ref{fig:unitary} to simplify this expression.
The step shown in Eq.~\eqref{eq:Bell_expect} immediately follows from exhausting all of the unitary contractions. 
This simplification can be seen as a statement of causality, such that gates outside of the observable light cone do not affect the expectation value.

\begin{figure*}
    \centering
    \includegraphics[width=\columnwidth]{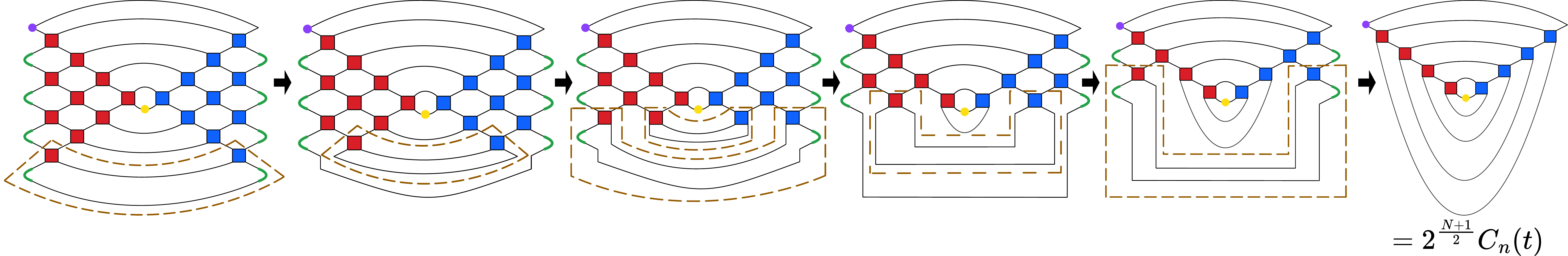}
    \caption[]{\textbf{Simplifying the expectation value through dual unitary contractions.} We start from the final expression of Eq.~\eqref{eq:Bell_expect}. 
    Dual-unitary contractions as indicated by dashed lines simplify the light cone structure on the left to the expression on the right which only includes gates on the light cone boundary.}
    \label{fig:Bell_proof_2}
\end{figure*}

To further simplify the expression in Eq.~\eqref{eq:Bell_expect}, we employ dual unitary contractions as shown in Fig.~\ref{fig:Bell_proof_2}.
We note that the final expression we arrive at is identical to the one given in Ref.~\cite{Ber-19a}, where identity (i.e., infinite temperature) initial states are used in place of our Bell pairs. 
Hence our expectation value is equivalent to the infinite-temperature autocorrelator $C_n(t)$.
Using similar contractions, it is easy to show that the results also match between the two different initial states in the case when $\hat{X_n}$ is not placed on the boundary of the lightcone of the $0$-th qubit -- both evaluate to zero.

With the end result of Fig.~\ref{fig:Bell_proof_2}, we can evaluate the tensor diagram by defining the following map on the local operator space: 
\begin{equation} 
\mathcal{M}_U[\hat{a}] = \quad \frac{1}{2}\adjincludegraphics[valign=c, width=0.2\linewidth]{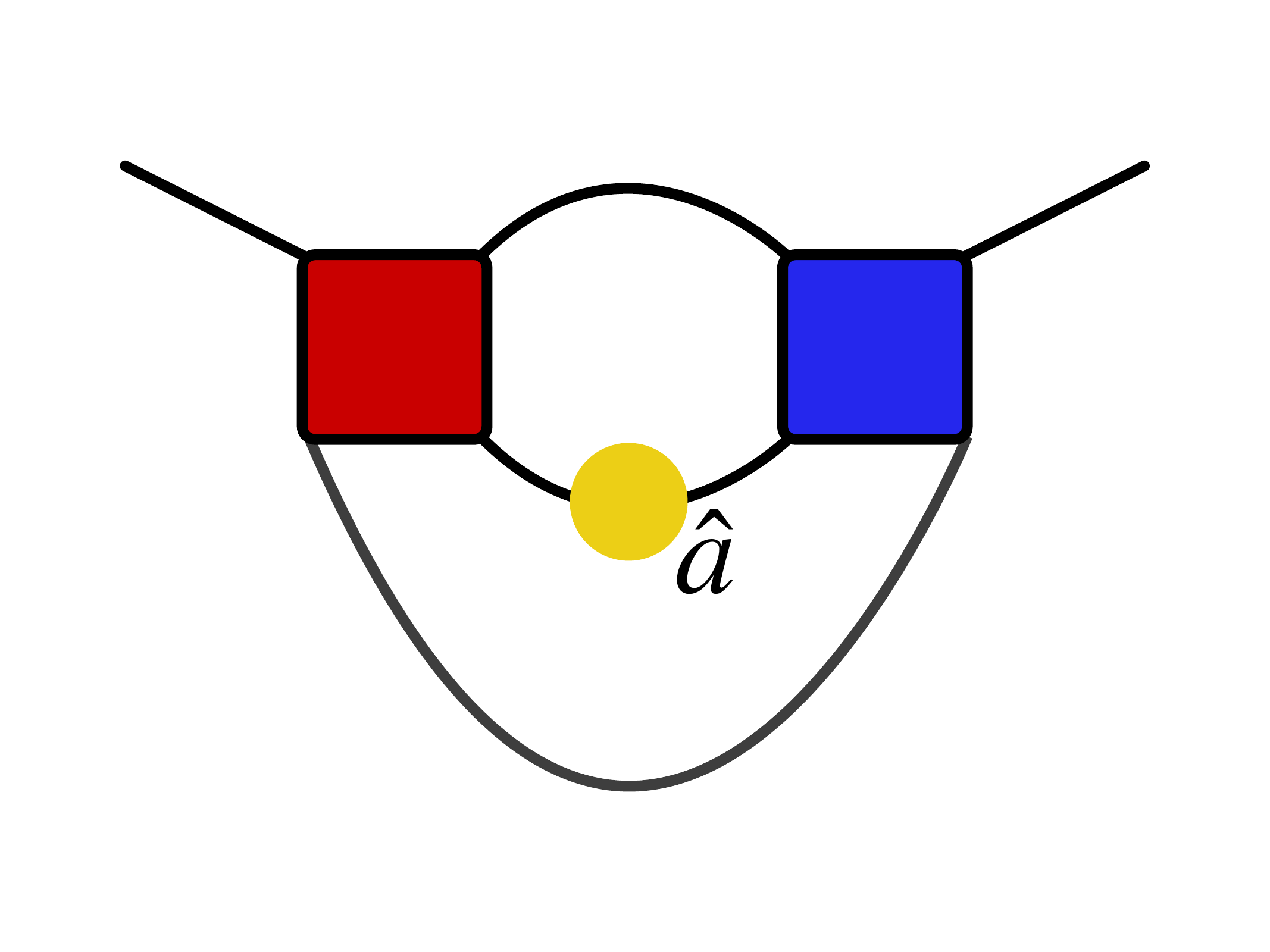} \quad = \frac{1}{2}\Tr_2\left(\hat{U}^\dagger(\hat{\mathbb{1}}\otimes \hat{a}) \hat{U}\right). 
\end{equation}
Keeping in mind that $|+\rangle\langle +| = (\hat{X} + \hat{\mathbb{1}})/2$ and $\Tr(M^k_U[\hat{X}]) = 0$ for any $k=1, 2, ...$, we may interchange $|+\rangle\langle +|$ and $\hat{X}$ in the role of $\hat{a}$ above provided we keep track of the factor of $2$.
Hence the correlator can be written as 
\begin{equation} C_n(t) = \Tr\left(|+\rangle\langle +|\mathcal{M}_U^n[\hat{X}]\right).
\end{equation}
Given the two-qubit unitary $U = e^{-ih\hat{Z}\otimes\hat{\mathbb{1}}} e^{-iJ\hat{Z}\otimes\hat{Z}} e^{-ib(\hat{X}\otimes\hat{\mathbb{1}}+ \hat{\mathbb{1}}\otimes\hat{X}) } e^{-iJ\hat{Z}\otimes\hat{Z}} e^{-ih\hat{Z}\otimes\hat{\mathbb{1}}}$ (see Eq.~\eqref{eq:two_qubit_gate}), we construct the matrix form of $\mathcal{M}_U$ in the Pauli basis $\{\hat{\mathbb{1}}, \hat{X}, \hat{Y}, \hat{Z}\}$: \begin{equation}
\mathcal{M}_U = \begin{bmatrix}
        1 & 0 & 0 & 0\\
        0 & \cos(2h) & 0 & 0\\
        0 & \sin(2h) & 0 & 0\\
        0 & 0 & 0 & 0
    \end{bmatrix}.\end{equation}
We finally arrive at the following result along the boundary of the lightcone (when $n=t$): \begin{equation} C_n(t) = \cos^t(2h).
\end{equation}

\section{Details on experiments}

\subsection{Device properties}

\begin{figure*}[b]
    \centering
    \includegraphics[width=\columnwidth]{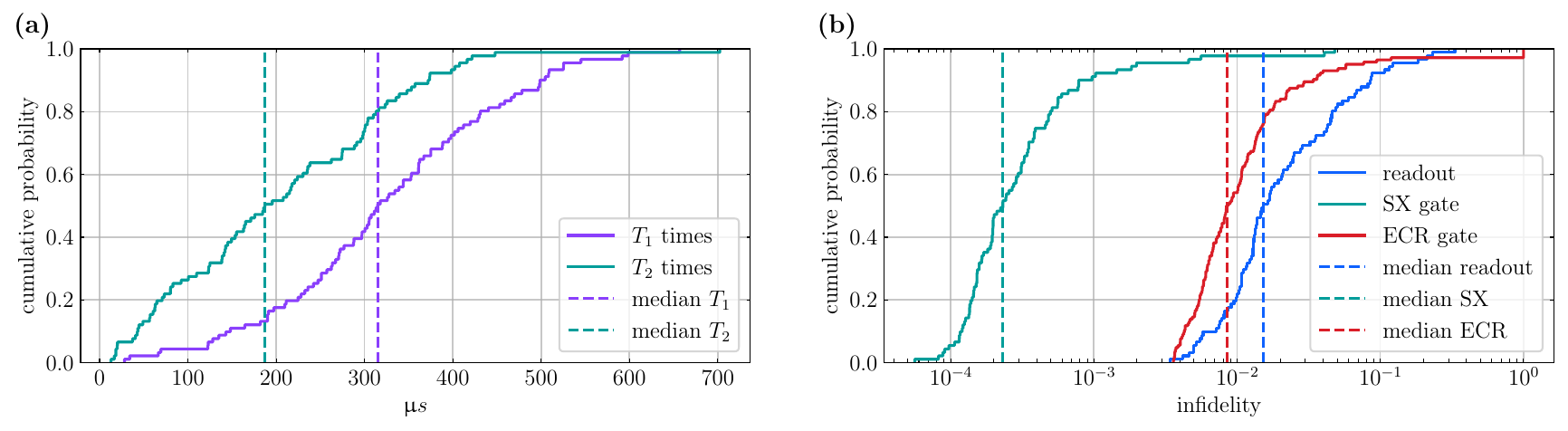}
    \caption[]{\textbf{Coherence times and error rates of} \textit{ibm\_strasbourg}. (a) Cumulative distribution of $T_1$ and $T_2$ times with median values.
    (b) Cumulative distribution of error rates for single-qubit $\sqrt{X}$-pulses (SX), two-qubit echoed cross-resonance (ECR) gates, and single-qubit readout, with median values indicated.}
    \label{fig:device_properties}
\end{figure*}

All experiments presented throughout this work are performed on the IBM Quantum Eagle processor \textit{ibm\_strasbourg} through a cloud-based access. 
The device consists of 127 fixed-frequency transmon qubits arranged in a heavy-hexagonal lattice (see qubit layout in Fig.~\ref{fig:device_noise_model}). 
We achieve a median $T_1$ time of $315\,\upmu s$ and median $T_2$ time of $187\,\upmu s$, see Fig.~\ref{fig:device_properties}(a) for the full distribution across the device.  
All the quantum circuits that we execute are decomposed into layers of parallel single-qubit gates and layers of parallel two-qubit entangling echoed cross-resonance (ECR) gates~\cite{sheldon2016procedure}. 
Single-qubit gates are implemented by $\sqrt{X}$-pulses (SX) and virtual $R_Z$ gates~\cite{mckay2017efficient}.
The distributions of the gate infidelities for the SX and ECR gates and the readout infidelities are shown in Fig.~\ref{fig:device_properties}(b). 
The median infidelities are $2.31\times10^{-4}$ for the SX gate, $8.53\times10^{-3}$ for the ECR gate, and $1.53\times10^{-2}$ for readout. 

\subsection{Quantum circuit execution}
\label{sec:circuit_execution}
The quantum circuits we consider in this work represent Floquet evolutions of a one-dimensional kicked Ising model with a Bell pair initialisation, see Sec.~\ref{sec:theory}.
These circuits consist of alternating layers of parallel ``odd'' layers ($1, 3, \dots $) and ``even'' layers ($0, 2, 4, \dots $) shown in Eq.~\eqref{eq:brickwork_timestep}.
When decomposed into the native gate set of the quantum processor, each two-qubit block of Eq.~\eqref{eq:brickwork_timestep} is transpiled into a sequence that includes two entangling ECR gates. 
Thus, each odd (even) step is built from two entangling layers of parallel ECR gates on odd (even) neighbouring qubit pairs, interleaved with layers of single-qubit gates.
In this Section, we will hence label the two unique ECR layers as ``even'' (for even time steps) and ``odd'' (for odd time steps as well as in the Bell pair initialisation). 
The executed circuits differ in their system parameters $J, b, h$ and the number of simulated time steps $t$, see Tab.~\ref{tab:circuit_parameter_settings} for an overview of all considered parameter settings.
We run all experiments at the scale of 51, 71 and 91 qubits. 
Before choosing the physical qubit layout, we perform calibrations of the single-qubit state preparation and measurement (SPAM) fidelities, $T_1$ times, and two-qubit Bell-pair preparation fidelities.
We then select 1d-chains of qubits that avoid outliers in these metrics.
For every model parameter set, we run several instances of the circuit while randomising measurements over the $\hat X$, $\hat Y$, and $\hat Z$ bases. 
This way, we obtain \emph{informationally complete}(IC) data as required for our error mitigation strategy, see Sec.~\ref{sec:informationally_complete_meas}.

We perform uniform Pauli twirling of the ECR gate layers to suppress coherent errors and obtain a Pauli noise channel~\cite{twirling_bennet, twirling_knill, wallman2016noise}. 
That is, for every parameter set, we run several instances of circuits that implement the same global unitary but differ in their single-qubit gate layers.
Similarly, we twirl measurements by inserting a Pauli $\hat X$ or $\hat I$ gate (sampled uniformly at random) prior to the readout, which we correct for in post-processing. 
This symmetrises the noise channel of the readout~\cite{van2022model}.

If done naively, gate twirling and randomised measurements create a circuit compilation overhead which can become prohibitively large for high circuit volumes and number of twirls. 
We alleviate this overhead by leveraging a recently introduced parametric circuit compilation and parameter binding pipeline facilitated by the \texttt{Sampler} primitive within the IBM Qiskit runtime service~\cite{qiskit2024}.
Each sequence of consecutive single-qubit gates on a given qubit (originating from the circuit itself, twirling, or readout basis rotation) is merged and implemented on the device with a sequence $R_z(\theta_3) \times \text{SX} \times R_z(\theta_2) \times \text{SX} \times R_z({\theta_1})$, parametrised by three angles $\theta_i$.
Hence, the twirled and randomised circuits are instances of the same parametrised circuit template and only differ in their $\theta$ angles.
With parametric compilation, we only need to create this underlying template circuit once, alongside the array of angles $\theta$ that represent the different twirled instances of the circuit. 
Our computational pipeline is thus significantly more efficient (both in terms of memory and execution time) than building the full circuit anew for every set of angles.
Nonetheless, the cost of resampling twirling and measurement configurations remains non-negligible, which is why we opt to collect multiple shots per setting. 
We take 1024 shots each for 256 randomised circuits per model parameter settings for a total of 262,144 shots per data point, see Tab.~\ref{tab:circuit_parameter_settings}. 
Error bars shown for unmitigated experimental data in Figs.~2 and~3 of the main text indicate one standard error across all individual shots.
The effect of repeated shots in the same measurement bases on the statistical errors is further discussed in Sec.~\ref{sec:postprocessing_outcomes}. 
In this way, we achieve a sampling rate ranging from $2.1\,\text{kHZ}$ (51-qubit dataset) to $1.57\,\text{kHZ}$ (91-qubit dataset). 

\begin{table}
\begin{center}
\begin{tabular}{|c||c|c|c|c|c|c|c|} 
 \hline
$N_\text{qubits}$ & \makecell{dual unitary \\ $J = b =\pi/4$ \\$ h = \{0, 0.05, 0.1, 0.15\}$} & \makecell{non dual unitary \\ $J=\pi/4, t= N_\text{qubits}$ \\$ h = \{0, 0.05, 0.1, 0.15\}$} &  \makecell{ circuit \\ randomisations} & \makecell{shots \\ per twirl} & \makecell{$R_z(\theta)$ gates \\at max. depth} & \makecell{wall clock \\time} & \makecell{sampling rate \\ in kHz} \\ 
 \hline\hline
 51 & $ t = \{0, 5, 10,  15, 20, 25\}$ & \multirow{3}{10em}{$b -\frac{\pi}{4} = \{ -0.15, -0.1, \dots, 0.15\}$} & \multirow{3}{1.5em}{256} & \multirow{3}{2em}{1024} & 7956 & 2h 18min & 2.10 \\ 
 71 & $ t = \{0, 7, 14, 21, 28, 35\}$ & & & & 15336  & 2h 55min & 1.75 \\
 91 & $ t = \{0, 9, 18, 27, 36, 45\}$ & & & & 25116 & 3h 24min & 1.57\\
 \hline
\end{tabular}
\caption{\textbf{Summary of parameter settings for quantum hardware execution.} 
The number of twirls and measurements (``shots'') per twirl are implemented for each combination of the model parameters $\{J, b, h, t\}$. 
The circuit randomisations include both gate twirling and the sampling of (twirled) randomised readout bases. 
The reported wall clock time is the total execution time for each dataset including readout error mitigation circuits (see Sec.~\ref{sec:readout_error_mitigation}), noise learning circuits (see Sec.~\ref{sec:noise_learning}), and additional benchmark circuits (see Sec.~\ref{sec:additional_circuits}).
The sampling rate is the total number of shots taken in each dataset divided by the wall clock run time.}
\label{tab:circuit_parameter_settings}
\end{center}
\end{table}

\subsection{Readout error mitigation}
\label{sec:readout_error_mitigation}

For the target kicked Ising circuits, our tensor network post-processing yields single-qubit observables $\langle \hat X_i\rangle^\text{TEM}$ where gate noise has been mitigated. However, these results are, in general, still affected by imperfect qubit readout. This effect can be removed by standard readout error mitigation techniques~\cite{maciejewski2020mitigation,geller2020rigorous}. In particular, we use a version of simple twirled readout error extinction (TREX) technique~\cite{van2022model}.
First, we characterise the state preparation and measurement (SPAM) error for all qubits by measuring the single-qubit $\langle 0|\hat  Z_i|0 \rangle$ expectation values while twirling the readout.
Here, we leverage the efficient parameter bindings provided by the control software stack, see Sec.~\ref{sec:circuit_execution}.
Then, we divide the expectation values $\langle \hat  X_i\rangle^\text{TEM}$ by the measured $\langle 0|\hat  Z_i|0 \rangle$ value of the corresponding qubit, thus obtaining the fully mitigated outcome.
The SPAM calibration circuits are interleaved with the kicked Ising circuits on a single-shot basis. 
In this way, the obtained $\langle 0|\hat  Z_i|0 \rangle$ value accurately matches the averaged SPAM value over the time interval for which the kicked Ising data is collected. 

\subsection{Noise learning}
\label{sec:noise_learning}

\begin{figure}
    \centering
    \includegraphics[width=0.99\columnwidth]{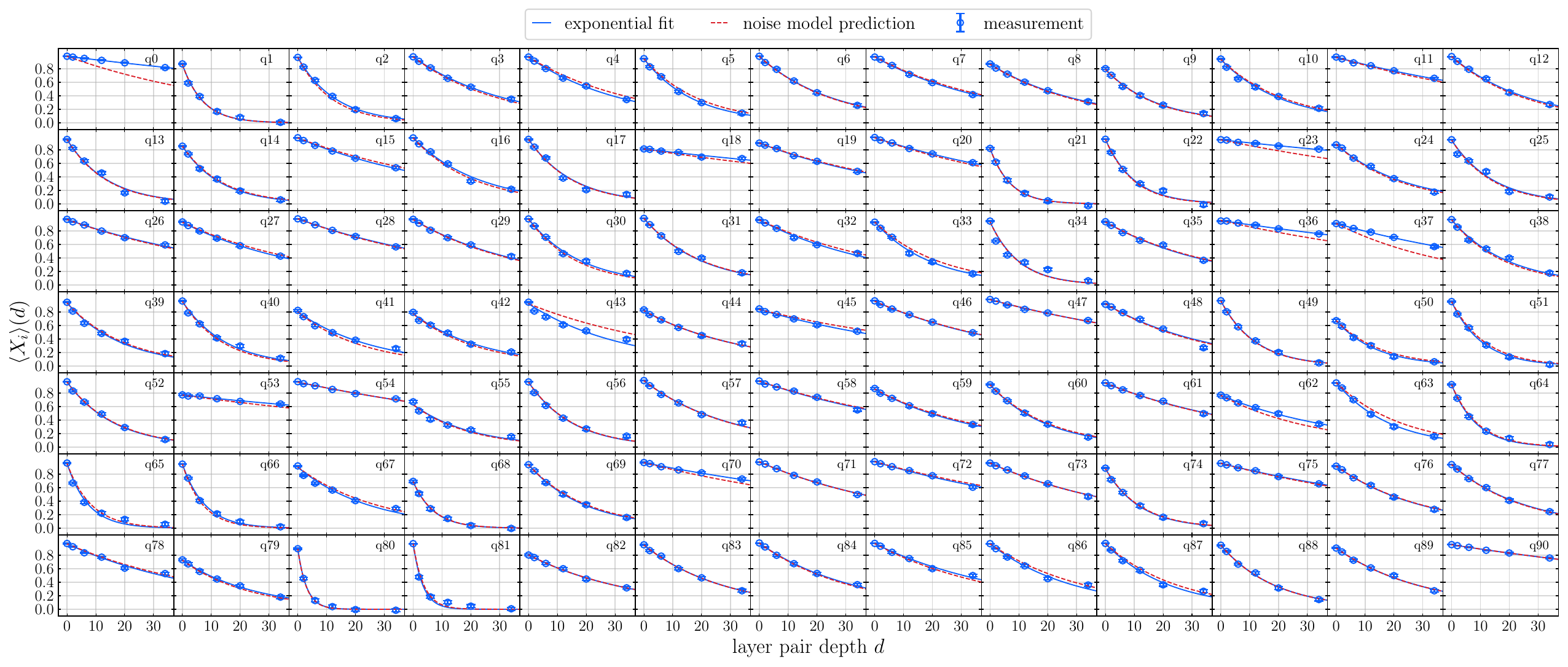}
    \caption[]{\textbf{Raw data for layer pair fidelity measurements.} Data points show the raw measured $\langle \hat X_i \rangle$ values for each qubit of odd ECR layer of the 91-qubit dataset with increasing layer pair depth $d$. 
    An exponential fit (solid lines) is applied to extract the single-qubit $X$ pair fidelities. For comparison, we show the prediction given by the obtained noise model (dashed lines) which matches the exponential fits well for most qubits. }
    \label{fig:CB_exponential_decays}
\end{figure}

\begin{figure*}
    \centering
    \includegraphics[width=\columnwidth]{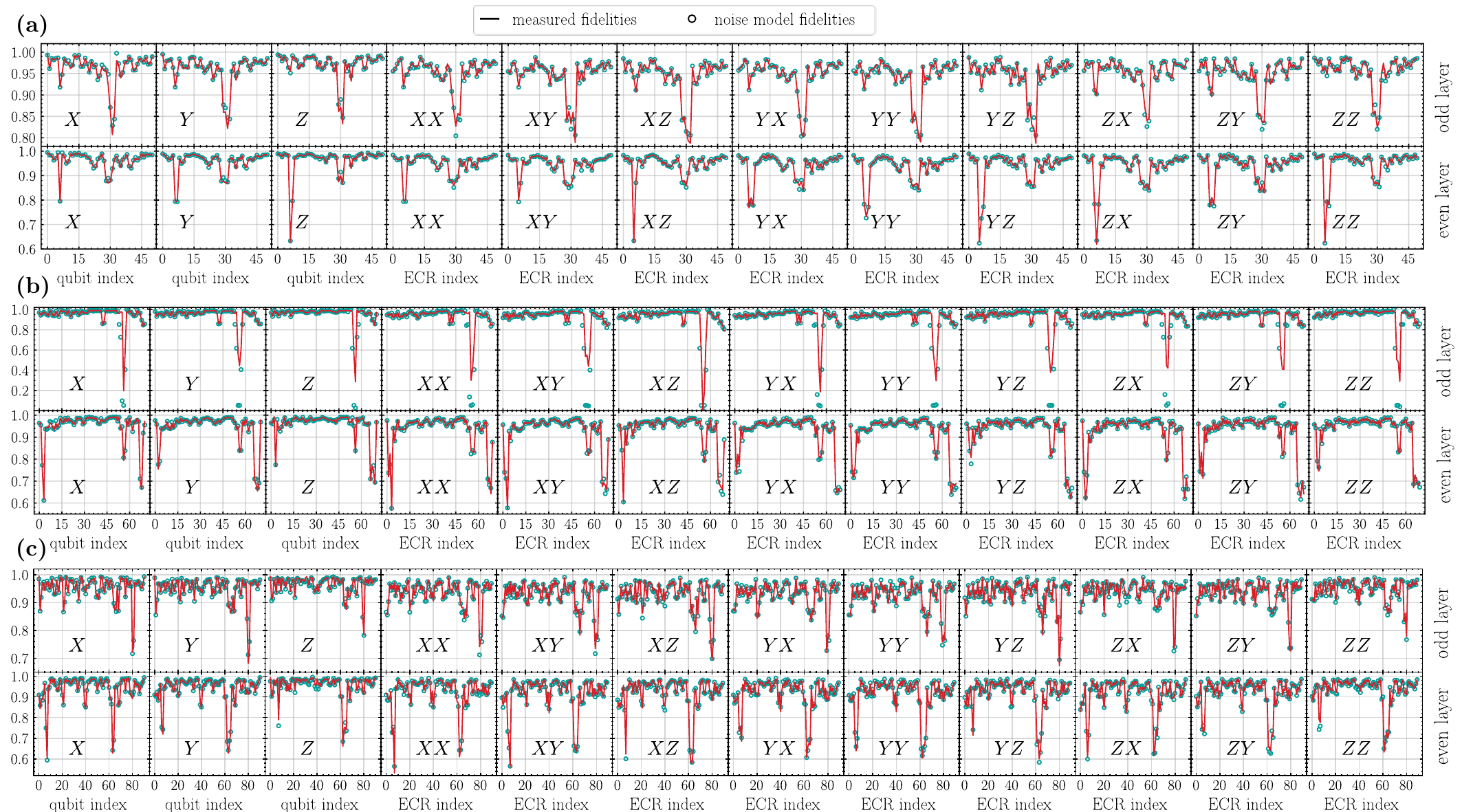}
    \caption[]{
    \textbf{Comparison of measured pair fidelities to the learned noise model.}
    For every dataset -- 51 qubits in (a), 71 qubits in (b), and 91 qubits in (c) -- the solid line indicates the measured single-qubit and two-qubit pair fidelities of each indicated Pauli of the sparse basis. 
    Circles show the value of that pair fidelity according to the fitted noise model. 
    }
    \label{fig:fidelities_comparison}
\end{figure*}

Our quantum circuits consist of two unique entangling ECR layers (labelled ``odd'' and ``even'') for which we accurately calibrate the noise in order to mitigate it.
We assume that the noise channel associated with the respective layer is identical whenever the layer appears in the circuit.
Let us denote the ideal unitaries of a layer as $U$ and the associated error channels as $\Lambda$, which are modelled to act before the unitaries.
For both ECR layers, we characterise $\Lambda$ building on the techniques developed in Ref.~\cite{van2023probabilistic}. 
For simpler notation, we drop hats on operators in the following.

Let $N_q$ be the number of qubits and $\mathcal{P}$ be the set of single-qubit and nearest-neighbour two-qubit Pauli operators $\mathcal{P} = \{\sigma_j \sigma_{j+1} \mid  \sigma_j, \sigma_{j+1} \in \{ I, X, Y, Z\}, j \in \{1, \dots, N_q \}\} $. The noise channels $\Lambda_{\text{odd}/\text{even}}$ are modelled as sparse Pauli-Lindblad channels of the form 
$\Lambda=\mathrm{e}^{\mathcal{L}}$ with
\begin{equation}
\label{eq:noise_model_def_supp}
\mathcal{L}(\rho) =  \sum_{P_i \in \mathcal{P}} \lambda_{i} \left( P_i \rho P_i^\dagger - \rho \right)
\end{equation}
parametrised by the generator rates $\lambda_{i}$.
Our task is to characterise the rates $\lambda_i$, which we obtain by fitting the sparse Pauli-Lindblad model to fidelities obtained from cycle benchmarking circuits~\cite{erhard2019characterizing, chen2023learnability}. 
In these circuits, we first prepare a $+1$ eigenstate of a given Pauli from $P_i \in \mathcal{P}$, then apply a given ECR layer $2d$ times, and finally measure $\langle P_i \rangle$.
As the ECR gate is self-inverse, every pair of layers in theory applies the identity operator.
The ideal measured expectation values should thus remain at $+1$, while, in practice, they decay due to noise. 
The ECR layers are also Clifford operations, so under conjugation with the layer $U$ a Pauli $P_i$ turns into a new Pauli 
\begin{equation}
\label{eq:definition_pauli_conjugate}
P_{i^\prime} = U P_{i} U^\dagger,
\end{equation}
which we also refer to as the conjugate Pauli of $P_i$.
We further define the Pauli fidelity of $P_i$ as 
\begin{equation}
\label{eq:Pauli_fidelity_def}
f_i = \Tr{\left( P_i {\Lambda} \left( P_i \right) \right)} / 2^n.
\end{equation}
Eqs.~\eqref{eq:definition_pauli_conjugate} and~\eqref{eq:Pauli_fidelity_def} imply that the decaying signal of the experiment is given by 
\begin{equation}
\label{eq:pair_fidelity_decay}
\langle P_i(d) \rangle =  (f_i f_{i^\prime})^d \times f^{\textsc{SPAM}}_i.
\end{equation}
where $f^{\textsc{SPAM}}_i$ is the state preparation and measurement fidelity associated with the prepared eigenstate.
We measure $\langle P_i(d) \rangle $ for various depths $d \in \{0, 2, 6, 12, 20, 34\}$ and perform an exponential fit to retrieve the \emph{pair fidelities} $\overline{f_i} := \sqrt{f_{i} f_{i^\prime}}$.
As an example of this, we show the measured decays of the single-qubit $X$ fidelities for the 91 qubit dataset in Fig.~\ref{fig:CB_exponential_decays}. 

So far, we have assumed that the error channels are Pauli channels. In reality, the noise maps acting on the device take a more general form, and include, for instance, non-unital terms and coherent gate imperfections. 
Such actual noise channels can, however, be shaped into Pauli form by performing Pauli twirling of the ECR gate layers, see Sec.~\ref{sec:circuit_execution}.
For the noise learning circuits, we sample 64 twirling instances for every depth $d$ with 32 shots per instance. 
For every depth $d$, different eigenstate initialisations, measurement bases, and twirling parameters then merely correspond to an updated set of $R_z(\theta)$ gate angles for the same parametrised circuit template.  
By measuring non-overlapping Paulis in parallel, a total of 9 different initial state settings are sufficient to cover the sparse basis $\mathcal{P}$~\cite{van2023probabilistic}.
The obtained pair fidelities $\overline{f_i}$ for all datasets presented in this work are shown in Fig.~\ref{fig:fidelities_comparison}.

\begin{figure}
    \centering
    \includegraphics[width=0.99\columnwidth]{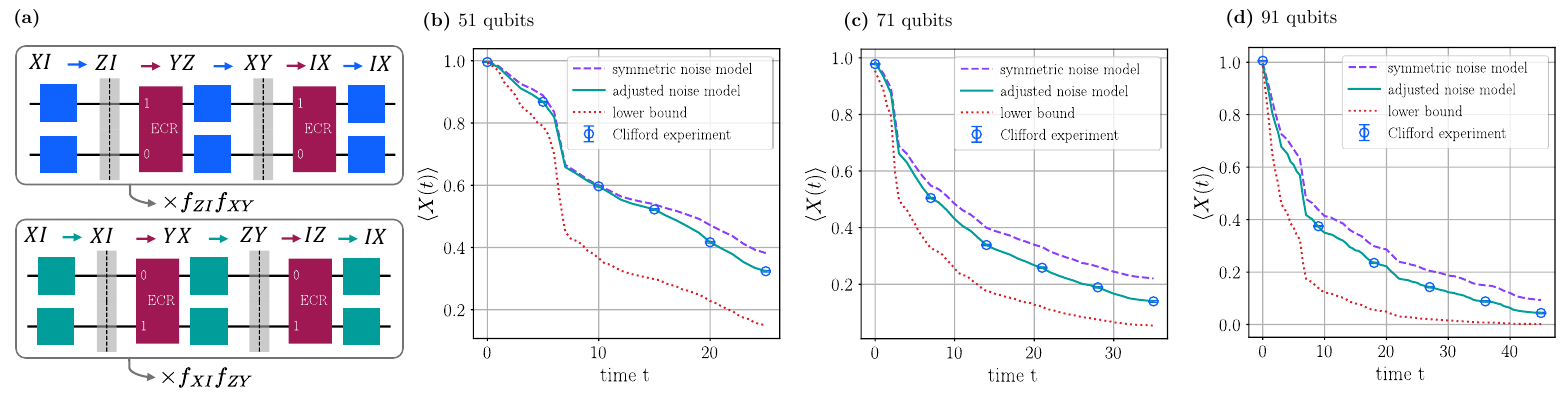}
    \caption[]{\textbf{Fine-tuning of the noise model with kicked Ising circuits at the Clifford point.}
    (a) The single-qubit $\langle X \rangle$ observable is affected by specific Pauli fidelities as the signal propagates through the noisy Clifford circuit. Here we show the contributing fidelities that depend on the direction of the ECR gate within the relevant two-qubit blocks (negative signs are omitted). 
    Note our convention of the noise occurring before the ECR gates, as represented by dashed lines. 
    The contributing single-qubit (two-qubit) Pauli fidelities are turned into two-qubit (single-qubit) fidelities and are thus not learnable in isolation by standard protocols. \\
    (b) -- (d) The noisy signal of the circuit is sensitive to the symmetry assumption between a given Pauli fidelity and its conjugate. 
    Traditionally, a symmetric split between these is assumed which predicts values (dashed lines) that do not match our experiments (round markers). 
    We thus adjust the underlying degrees of freedom to obtain a noise model that matches the experiment (solid lines). 
    This is well within the region of physically allowed values indicated by the lower bound (dotted lines).
    }
    \label{fig:tweaked_fidelities_with_ki}
\end{figure}

According to Eq.~\eqref{eq:pair_fidelity_decay}, our protocol does not distinguish between the fidelity of a given Pauli and its conjugate and thus only learns self-conjugate fidelities reliably. 
In previous error mitigation works the generators $\lambda_i$ were obtained under the symmetry assumption that $f_i = f_{i^\prime}$~\cite{van2023probabilistic, kim2023evidence}. 
In principle, it is known that fidelities of Pauli operators whose conjugate Pauli has the same Pauli weight can be learned with an ``interleaved'' cycle benchmarking protocol~\cite{chen2023learnability}. 
However, the fidelities of Pauli operators that do change weight under conjugation with the layer $U$ remain fundamentally unlearnable in a SPAM-robust way.

Let us now discuss how this limitation affects the kicked Ising Floquet circuits.  
In Fig.~\ref{fig:tweaked_fidelities_with_ki}(a) we show the propagation of Paulis for the desired $X$-observable of the two-qubit dual unitary circuit blocks at the Clifford point. 
Depending on the control-target direction of the \textsc{ECR} gate, the observable is affected by a factor of $f_{ZI} f_{XY}$ or $f_{XI} f_{ZY}$. 
However, the conjugate fidelities of these are $f_{ZI^\prime} = f_{YZ}$, $f_{XY^\prime} = f_{IX}$, $f_{XI^\prime} = f_{YX}$, and $f_{ZY^\prime} = f_{IZ}$. 
Hence the Clifford signal is a product of fidelities that can not be individually learned by standard cycle benchmarking circuits. 
As a result, the kicked Ising circuits are highly sensitive to the underlying symmetry assumptions of the noise model.

\begin{figure*}
    \centering
    \includegraphics[width=\columnwidth]{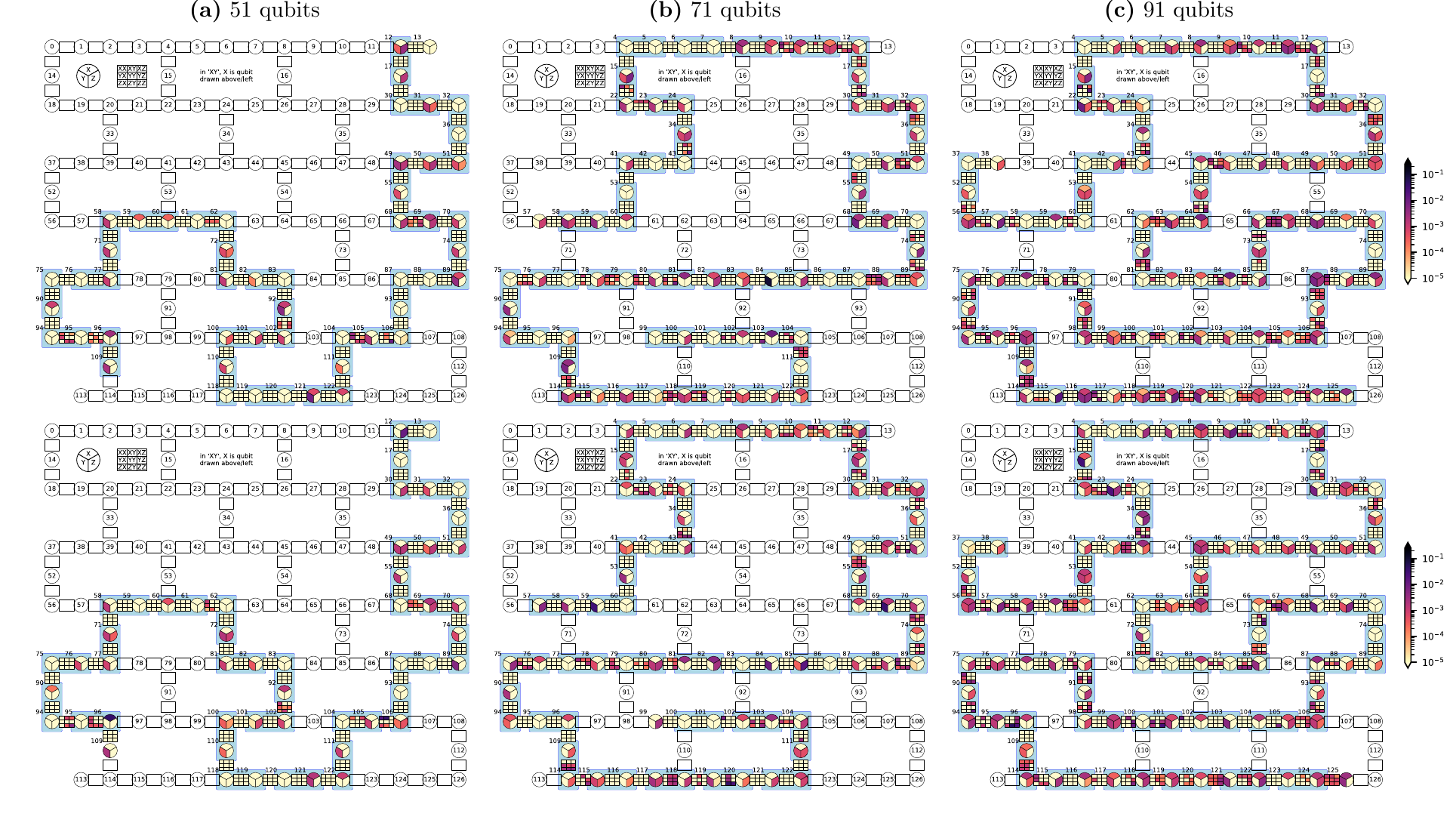}
    \caption[]{\textbf{Device noise models of ECR layers for selected 1d-chains on} \textit{ibm\_strasbourg}. 
    The device layout is a heavy-hexagonal lattice where qubits are represented by circles and rectangles denote qubit connections. Error generators $\lambda_i$ of all single-qubit terms and nearest-neighbour two-qubit terms are indicated by he colour scale. 
    The top row shows the noise models of the odd ECR layer, while the bottom row shows the even layer. 
    Blue boxes indicate the pairs of qubits between which ECR gates are implemented.
    }
    \label{fig:device_noise_model}
\end{figure*}

Fig.~\ref{fig:tweaked_fidelities_with_ki}(b) -- (d) shows the comparison of the Clifford point experiment (after readout error mitigation, see Sec.~\ref{sec:readout_error_mitigation} for details) with the measured pair fidelities. 
Indeed, the product of characterised pair fidelities deviates from the measured values for $\langle X(t) \rangle$ indicating that noise models based on the symmetry assumption do not reflect the noise of the device with sufficient accuracy. 
This motivates us to introduce asymmetric weights $\alpha_i$ for every fidelity $f_i$ that contributes to the $\langle X(t) \rangle$ signal at the Clifford point. 
We define re-weighted fidelities as $f_i(\alpha_i) = \alpha_i \overline{f_i}$ and $f_{i^\prime}(\alpha_i) = \overline{f_i} / \alpha_i$.
This way, the pair fidelities $\sqrt{f_i(\alpha_i) f_{i^\prime}(\alpha_i)}$ remain independent of $\alpha_i$.
Our goal is to find parameters $\alpha_i$ such that the product of the relevant fidelities $\prod_i f_i(\alpha_i)$ matches the measured value in the kicked Ising experiment at the Clifford point. 
For a physical (positive and trace-preserving) channel $\Lambda$, Pauli fidelities are bounded as $f_i \leq 1$. 
To ensure this, $ \alpha_i^\text{min} \leq \alpha_i \leq \alpha_i^\text{max}$ with $\alpha_i^\text{min} = \overline{f^C_i}$ and $\alpha_i^\text{max} = 1/\overline{f^C_i}$ must hold. 
We can thus more effectively parametrise the weighted fidelities by $\delta_i \in \left[ 0, 1 \right]$ as $\alpha_i (\delta_i) = \delta_i \alpha_i^\text{min} + (1-\delta_i) \alpha_i^\text{max}$. 

Since there are more relevant weights $\alpha_i$ than available data points of the Clifford observable, their choice when fitting the noise model to the experiment values is not unique. 
Our procedure for obtaining $\alpha_i$ is then the following: 
starting from the Clifford data point at lowest depth $>0$, we choose a uniform $\delta$ for all $\delta_i$ that affect the data point, such that the noise model prediction matches that value. 
We then iteratively move to the next data point, choosing a new uniform $\delta$ for the fidelities that enter the signal between that and the previous data point, until all data points are in agreement with the chosen fidelity splits.
This procedure ensures that the resulting values for $\alpha_i$ avoid edge cases where one of the fidelities becomes $\approx 1$. 
For those fidelities that are not probed by the Clifford experiments, we continue to assume a symmetric split ($\alpha_i = 1$).

The question arises of how exhaustively we must exploit the range of possible values for $\alpha_i$ to match the Clifford experiment. 
We show the lower bound on the measured observable obtained from $\prod f^C_i(\alpha_i^\text{min})$ as a dotted red line in Fig.~\ref{fig:tweaked_fidelities_with_ki}(b) -- (d). 
This confirms that the chosen fidelity splits are well within their allowed physical regions. 
However, we observe a consistent trend that the splits need to be chosen such that the contributing fidelities become lower (and the conjugate ones higher). 
This indicates that there either is a systematic physical mechanism that causes the fidelity splits to fall on this side or that certain noise sources are present that are not fully captured by the sparse Pauli-Lindblad model. 
This point is further investigated in Sec.~\ref{sec:additional_circuits}. 

Next, for both the odd and even layer, we fit the generator rates $\lambda_i$ of the noise model from Eq.~\eqref{eq:noise_model_def_supp} to the obtained Pauli fidelities.
Let $M$ be a square matrix with entries $M_{ij} = 1, i, j \in \mathcal{P}$ if $\{ P_i, P_j \} = 0$ and $M_{ij} = 0$ otherwise. 
Similarly, we define $M^\prime$ with entries  $M^\prime_{ij} = 1$ if $\{ U P_i U^\dagger, P_j \} = 0$ and $M^\prime_{ij} = 0$ otherwise.
We use $\boldsymbol{\lambda}$ to denote the vectors with entries of $\lambda_i$ (and similarly for $f_i$, $f_i^\prime$, $\overline{f}_i$ and $\alpha_i$). 
The relationship between generators and fidelities is then given as $ M \boldsymbol{\lambda} = \log(\boldsymbol{f})/2$ and $M^\prime \boldsymbol{\lambda} = \log(\boldsymbol{f^\prime})/2$. 
Building on Ref.~\cite{van2023probabilistic}, we find the generators that best describe the obtained fidelities by solving the non-negative least-squares problem 
\begin{equation}
\label{eq:generator_fidelities_fit_supp}
\boldsymbol{ \lambda}_\text{fit} :=  \operatorname*{arg\,min}_{\lambda_i \geq 0} \; \Bigg\lVert
\left[\begin{array}{c}M \\ M^\prime \end{array}\right] \boldsymbol{\lambda}
+ \frac{1}{2} \log \left[\begin{array}{c} \boldsymbol{f}(\boldsymbol{\alpha}) \\ \boldsymbol{f^\prime}(\boldsymbol{\alpha}) \end{array}\right]  \Bigg\rVert_2^2. 
\end{equation}

We can assess the validity of the noise model obtained in this way by comparing the predicted pair fidelities $\overline{\boldsymbol{f}} = \exp\left(2 ( M + M^\prime) \boldsymbol{\lambda}\right)$ (with element-wise exponentiation) to the measured pair fidelities in Fig.~\ref{fig:fidelities_comparison}.
Overall, the noise model is in good agreement with the measured pair fidelities. 
There are small deviations mostly around especially low fidelities (e.g., between qubits 50 -- 60 for the 71 qubit odd layer), suggesting the presence of noise contributions beyond the sparse Pauli-Lindblad model. 
The prediction of the obtained noise models for the Clifford point of the kicked Ising circuits is shown in Fig.~\ref{fig:tweaked_fidelities_with_ki}(b) -- (d) (solid green line). 
This confirms that our noise models are consistent with both the noise learning circuits as well as the kicked Ising Clifford circuits. 
The resulting generators $\lambda_i$ for all considered noise models are visualised in Fig.~\ref{fig:device_noise_model} where we also indicate the chosen physical qubit layout of each experiment. 

The above procedure can be regarded as a fine-tuning of the hitherto state-of-the-art noise learning pipeline to our particular application by treating Clifford circuits as additional learning circuits. 
At this stage, a natural question concerns how our apparent ability to fit individual fidelities can be reconciled with the unlearnability statements from Ref.~\cite{chen2023learnability} quoted above. 
The subtle resolution of this seeming contradiction is that our learning of the fidelity splits from the Clifford circuits is no longer independent of SPAM errors. 
By applying twirled readout error mitigation to the kicked Ising observables, we implicitly assume that only readout errors contribute to the SPAM and state preparation is essentially perfect. 
This assumed ``gauge'' puts constraints on the fidelities of the gate noise.
The splits of previously unresolved fidelity pairs could then be learned by suitably prepared depth-one circuits~\cite{chen2024efficientselfconsistentlearninggate}. 
Note, however, that the individual fidelities can not be amplified and are thus more difficult to estimate accurately. 
Our noise learning protocol is a first step in this direction. 

\subsection{Additional benchmark circuits}
\label{sec:additional_circuits}

\subsubsection{Repeated odd/even kicked Ising layers}
\label{sec:repeated_DU_layers}

\begin{figure*}
    \centering
    \includegraphics[width=\columnwidth]{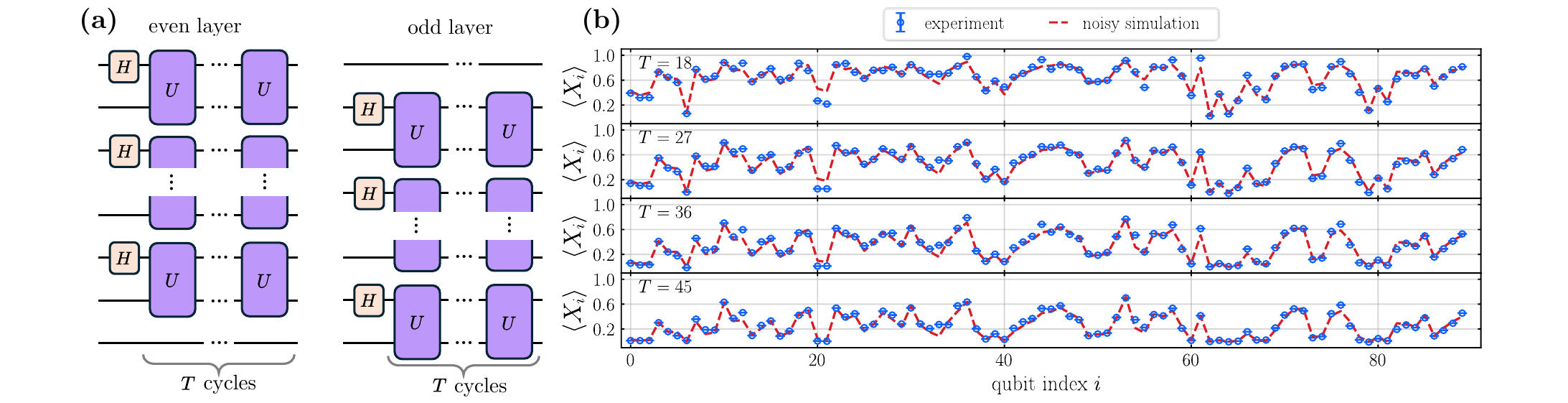}
    \caption[]{\textbf{Repeated odd/even kicked Ising layer benchmark experiments.} (a) Benchmark circuits consist of repeated layers of parallel two-qubit kicked Ising blocks at the Clifford point, applied on all even and odd qubit pairs for different cycle depths $T$, respectively. $H$ denotes the Hadamard gate.  
    (b) Measured $\langle X_i\rangle$ expectation values compared to a stabiliser simulation given the learned noise model. 
    Note that the ECR layer depth is $2T$.}
    \label{fig:repeated_DU_blocks}
\end{figure*}

In this Section, we further investigate how well the noise model generalises to different benchmark observables for circuits built from the same even and odd ECR layers for which the noise was characterised. 
First, we examine the two-qubit building blocks of the kicked Ising model as defined in Eq.~\eqref{eq:two_qubit_gate}.
Specifically, we run two sets of quantum circuits where we implement repetitions of the even dual unitary layer $\hat{\mathbb{U}}_e$ and the odd dual unitary layer $\hat{\mathbb{U}}_o$ at the Clifford point ($J=b=\pi/4, h=0$, respectively).
For the repeated even (odd) layer circuits, we initialise every even (odd) qubit in the $\ket{+}$ state, see Fig.~\ref{fig:repeated_DU_blocks}(a). 
After $T$ cycles of the repeated even (odd) layers, we measure the single-qubit $\langle X_i \rangle$ observable for every even (odd) qubit index $i$ if $T$ is even or for every odd (even) qubit index $i$ if $T$ is odd. 
In this way, we obtain $N-1$ expectation values $\langle X_i \rangle$, where $N$ is the number of qubits. 
As for the other kicked Ising experiments, we take 256 twirling randomisations of the circuit with 1024 shots per circuit. 

Fig.~\ref{fig:repeated_DU_blocks}(b) shows the measured $\langle X_i \rangle$ values of different depths $T$ for the 91-qubit data set, where readout error mitigation has been applied as described in Sec.~\ref{sec:readout_error_mitigation}. 
We compare these values to a noisy simulation of the circuits given the noise model learned for the ECR layers. 
Since the circuit consists of Clifford gates and the noise channels are Pauli error channels, we can simulate this efficiently in the stabiliser formalism by propagating the Heisenberg-evolved observable backwards through the circuit.
The prediction of the learned noise model indeed matches the experimental values well. 
The main difference of these circuits to the noise learning circuits is the additional single-qubit gates in between the ECR layers. 
This experiment can thus be seen as an interleaved cycle benchmarking run whose Pauli cycle consists of the particular pair fidelities that enter the two-qubit kicked Ising blocks when the $X$ operator propagates through it.
Our noise model predicts these decays well up to a depth of 22.5 cycles ($T=45$, ECR depth 90). 

\subsubsection{Mirror circuits of Floquet evolution}
\label{sec:mirror_circuits}
\begin{figure*}
    \centering
    \includegraphics[width=\columnwidth]{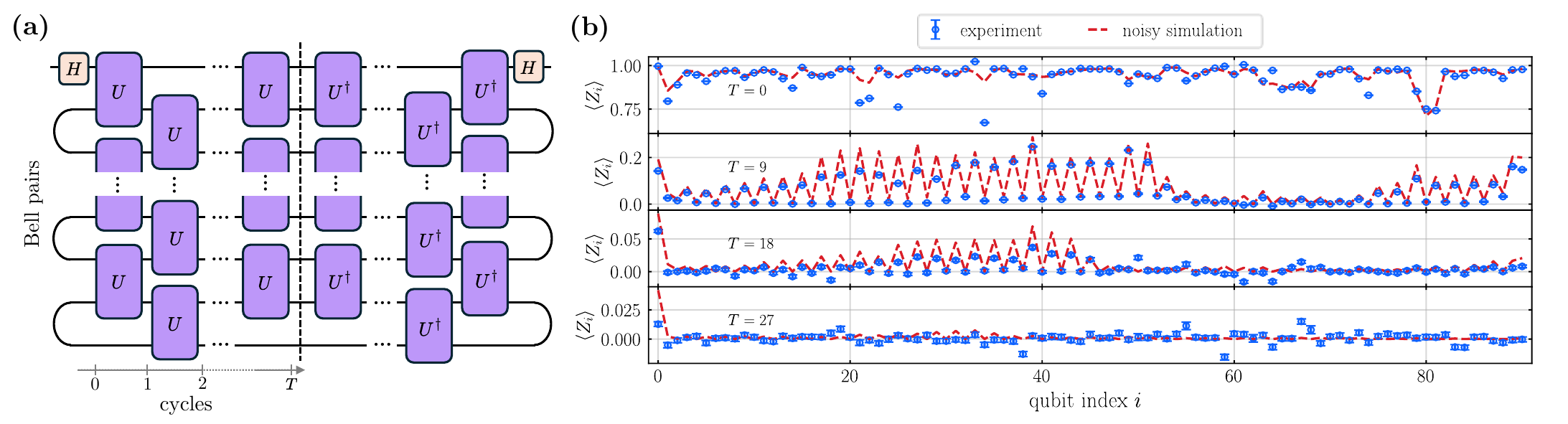}
    \caption[]{\textbf{Mirrored kicked Ising benchmark experiments.} (a) The circuits consist of the forward-time kicked Ising evolution for $T$ Floquet cycles with Bell pair initialisation at the Clifford point, followed by the inverse gates in reverse order to create a mirrored identity circuit. 
    (b) Measured $\langle Z_i\rangle$ expectation values compared to a stabiliser simulation given the learned noise model. 
    Note the different vertical scales across panels, and that the ECR layer depth is $4T + 2$.}
    \label{fig:mirror_circuits}
\end{figure*}

Another class of circuits we run is the Clifford point of the kicked Ising experiment followed by a mirrored ``uncomputation'' of the entire circuit, i.e., applying the inverse unitary of each gate in reverse order, see Fig.~\ref{fig:mirror_circuits}. 
These circuits thus first implement the forward-time evolution of the Floquet dynamics for some depth $T$, then run the reverse time evolution and finally undo the initial state preparation, ideally recovering the reference $\ket{0}^{\otimes N}$ state. 
Since the entangling layers are self-inverse, this circuit still only consists of the two unique ECR layers and single-qubit gates. 
We measure single-qubit $\langle Z_i \rangle$ expectation values for every qubit. 
Note that these observables are intrinsically insensitive to the pair fidelity weights $\boldsymbol{\alpha}$ used to fit the noise model in Eq.~\eqref{eq:generator_fidelities_fit_supp}.
This is because the uncomputation gates always pick up the conjugates of the fidelities that enter the noisy signal during the forward evolution. 
These circuits thus form a benchmark of the noise model that is independent of the assumptions on symmetry in Pauli noise learning (and SPAM mitigation). 

Fig.~\ref{fig:mirror_circuits}(b) shows the obtained $\langle Z_i \rangle$ expectation values (readout error mitigated) alongside a noisy Clifford simulation. 
For even qubit indices (except $i=0$) the values decay quickly with increasing $T$ as the Pauli weight of the contributing fidelities grows linearly in the forward evolution for these observables.
In contrast, for odd $i$, the Pauli weight of the contributing fidelities never grows beyond four, which explains the zig-zag shapes of the measured values. 
This pattern is qualitatively well reflected in the measured data. 
However, quantitatively, some of the measured expectation values fall below the prediction of the noise model. 
This indicates that our circuits are subject to small additional noise sources that are not entirely captured by our noise model. 

Interestingly, the repeated odd/even layer circuits from Sec.~\ref{sec:repeated_DU_layers} are not affected by an underestimation of noise. 
Indeed, in contrast to the $\langle Z_i \rangle$ observables of the mirror circuit, the Pauli fidelities that contribute to the noise of $\langle X_i \rangle$ in those circuits are confined to two-qubit strips (which also explains why they do not decay as quickly).
This is closer to the fidelity cycles measured in the noise learning circuits which extend at most to four-qubit strips. 
Our data thus suggests that -- when the Pauli weight pattern of the observable traverses larger regions of the qubit lattice -- there are additional noise sources that our noise model does not account for.
Candidates for this include higher-order or non-nearest-neighbour noise generators, residual coherent errors, as well as leakage, i.e., transmon states with population outside of the qubit subspace. 
We leave a more thorough investigation of these effects and their implications on error mitigation for future work. 

\subsection{Timing of experiments and stability of the noise model}

For noise learning based error mitigation techniques, temporal drifts of the device noise model may cause imperfections in the mitigated results.
The question arises if this effect may explain the slight mismatch between the noise model prediction and the observed expectation values for the mirrored circuits shown in Sec.~\ref{sec:mirror_circuits}. 
The order in which the experiments are run on the device is the following: We first run the circuits of repeated odd/even kicked Ising layers from Sec.~\ref{sec:repeated_DU_layers}, followed by the mirror circuit experiments. 
Next, we perform the main experiments of all kicked Ising circuits summarised in Tab.~\ref{tab:circuit_parameter_settings} and finally run the noise learning circuits as outlined in Sec.~\ref{sec:noise_learning}. 
Hence, the repeated odd/even kicked Ising layers are the circuits most separated from the noise learning circuits in time. 
Yet they match the prediction of the obtained noise model accurately, indicating that the noise model is stable in time over the duration of the experiments (up to 3.5h for the 91 qubit data set). 
We thus believe that the minor mismatch observed for the mirror circuits is a more systematic effect rather than caused by temporal drifts.
This is further corroborated by the fact that the observed mismatch systematically tends towards lower values (underestimation of the noise) rather than spreading into both directions, which we would expect from stochastic drifts in time.

\section{Classical Processing Procedures}

\subsection{Matrix product operators in the Pauli transfer matrix representation}
\label{sec:MPO_PTM}
\emph{Matrix product operators} (MPOs) are commonly used for the efficient representation and manipulation of linear operators that act on quantum systems~\cite{SCHOLLWOCK201196, PhysRevB.95.035129, Montangero2018}. In this work, we use MPOs in the \emph{Pauli transfer matrix} (PTM) representation, which reduces the action of quantum channels to matrix multiplication and makes channel inversion straightforward~\cite{Wood:2011zvw}.
In the PTM representation, quantum operators are described as linear combinations of Pauli matrices. For a single qubit, an operator $\mathcal{O}$ can be expressed as:
\begin{equation}
    \label{eq:ptm_representation}
    \mathcal{O} = \sum_{\alpha, \beta \in \{I, X, Y, Z\}} \mathcal{O}_{\alpha \beta} P_{\alpha }\otimes P_{\beta}
\end{equation}
where $P_{\alpha}$ and $P_{\beta}$ are Pauli matrices, and $\mathcal{O}_{\alpha,\beta}$ are the elements of the PTM. 
For an $N$-qubit system, the operator acts as a tensor product of Pauli matrices across all qubits:
\begin{equation}
    \label{eq:PTM_Nq}
    \mathcal{O} = \sum_{\Vec{\alpha}, \Vec{\beta}} \mathcal{O}_{\Vec{\alpha},\Vec{\beta}} \  P_{\Vec{\alpha}} \otimes P_{\Vec{\beta}}
\end{equation}
where $\Vec{\alpha}$ and $\Vec{\beta}$ are $N$-tuples representing the Pauli indices for each qubit and $\mathcal{O}_{\Vec{\alpha},\Vec{\beta}}$ is the corresponding PTM element.
To represent $\mathcal{O}$ as an MPO, we first decompose it into a product of local tensors $\mathcal{O}^{[q]}$ associated with each qubit $q$. Each tensor has the structure $\mathcal{O}^{[q]}_{(\alpha_q, \beta_q, \gamma_{q - 1}, \gamma_{q})}$ where $\alpha_k$ and $\beta_k$ are the physical indices of size $4$ representing the Pauli operators for qubit $q$ and $\gamma_{q - 1}$ and $\gamma_q$ are the bond indices that connect the tensors of adjacent qubits $q - 1$ and $q$. The full MPO is then written as 
\begin{equation}
    \label{eq:full_MPO}
    \mathcal{O} = \sum_{\gamma} \mathcal{O}^{[0]}_{\gamma_0} \otimes \mathcal{O}^{[1]}_{\gamma_0 \gamma_1} \otimes \dots \otimes \mathcal{O}^{[N - 2]}_{\gamma_{n - 3} \gamma_{n - 2}} \otimes \mathcal{O}^{[N - 1]}_{\gamma_{N - 2}}
\end{equation}
where, to simplify the notation, we omitted the physical indices which are attached to each tensor. In this form, the action of the channel MPO is simplified to matrix multiplication which is performed via contraction of the shared Pauli indices. Consider, for instance, two $N$-qubit MPOs in PTM representation, say $\mathcal{A}$ and $\mathcal{B}$ defined as 
\begin{align}
    \label{eq:mpo_contraction}
    \mathcal{A} = \sum_a \mathcal{A}_{a_0}^{[0]} \otimes \mathcal{A}_{a_0a_1}^{[1]}\otimes \dots \otimes \mathcal{A}_{a_{N - 2}}^{[N - 1]}\\
    \mathcal{B} = \sum_b \mathcal{B}_{b_0}^{[0]} \otimes \mathcal{B}_{b_0b_1}^{[1]}\otimes \dots \otimes \mathcal{B}_{b_{N - 2}}^{[N - 1]}.
\end{align}
If we compute $\mathcal{C} = \mathcal{A}\mathcal{B}$ we get a resulting MPO that represents the product operator in the PTM space
\begin{equation}
    \label{eq:multiplied_mpo}
    \mathcal{C} = \sum_c \mathcal{C}_{c_0}^{[0]} \otimes \mathcal{C}_{c_0c_1}^{[1]}\otimes \dots \otimes \mathcal{C}_{c_{N - 2}}^{[N - 1]}
\end{equation}
where $\mathcal{C}$ has bond indices $c_q$ that are a multi-index composed of virtual indices ($a_q, b_q$) with dimensions $|{c_q}| = |{a_q}|\cdot|{b_q}|$.

\subsection{Compression}
\label{sec:MPO_MPS_compression}
The multiplicative growth of the bond dimensions in the MPO induced by composition can lead, in general, to an exponential increase in the size of the relevant tensors. To manage computational resources efficiently, it is therefore necessary to design a suitable MPO compression procedure.
To reduce the bond dimension of an MPO, a common method is to truncate the singular values in the canonical form of the operator~\cite{dmrg_2007, PhysRevB.95.035129}. To do this, a \emph{singular value decomposition} (SVD) is performed on the tensors that define the MPO at each bond. For example, to truncate the first bond of the resulting MPO in Eq.~\eqref{eq:multiplied_mpo} let us consider the tensor located at the first site $\mathcal{C}^{[0]}_{(\alpha_0 , \beta_0, c_0)}$. 
To analyse only the shared bond index between qubit $0$ and $1$ we first reshape this tensor to have one multi-index combining the physical indices. 
$\mathcal{C}^{[0]}_{(\alpha_0 , \beta_0, c_0)} \rightarrow \mathcal{C}^{[0]}_{(\alpha_0 , \beta_0), c_0}$
Next, an SVD is performed to give $\mathcal{C}^{[0]}_{(\alpha_0 , \beta_0), c_0} = U^{[0]}_{(\alpha_0, \beta_0), \delta} S^{[0]}_{\delta} (V^{[0]})^{\dagger}_{\delta , c_0}$ where $U^{[0]}$ is a unitary matrix whose columns are the left singular vectors corresponding to the combined physical indices, $S^{[0]}$ is a diagonal matrix of singular values $\lambda_{\delta}$ and $V^{[0]}$ is another unitary matrix whose columns are the right singular vectors corresponding to the bond index $c_0$. After SVD, the singular values $\lambda_{\delta}$ are ordered from largest to smallest. The magnitude of these singular values indicates how much each corresponding mode (combination of physical and bond indices) contributes to the shared bond. A truncation in which the smallest singular values are discarded is then justified, as it corresponds to retaining only the most significant modes. We then modify $\mathcal{C}^{[0]'}_{(\alpha_0 , \beta_0, c_0)} = U^{[0]}_{(\alpha_0, \beta_0), \delta'} S^{[0]}_{\delta'}$ with $\delta' \leq \delta $ and propagate the truncated $V^{[0]}$ to the next tensor, giving $\mathcal{C}^{[1]'}_{(\alpha_1 , \beta_1, c_0, c_1)} = V_{\delta' , c_0}^{[0]} \mathcal{C}^{[1]}_{(\alpha_1 , \beta_1, c_0, c_1)}$. This is performed sequentially on all qubits, compressing each of the bonds such that the resulting MPO will, in general, have a smaller bond dimension, depending on the truncation procedure. Two different truncation approaches have been used in this work. For the classical simulations we choose \(\epsilon=10^{-12}\) as a cutoff in the truncation procedure: this means that after each SVD in the compression routine we keep at most \(m\) singular values, where \(m\) is the minimum between the maximum allowed bond dimension and the smallest \(k\in\{1,\dotsc,M\}\) such that
\begin{equation}
  \frac{\sum_{j=k+1}^{M}\lambda_j^2}{\sum_{j=1}^{M}\lambda_j^2}<\epsilon.
\end{equation}
Here, we assume that the singular values \(\{\lambda_j\}_{j=1}^{M}\) are given in decreasing order.
This ensures that the total error (i.e.~the 2-norm distance between the original and the compressed MPO) will not be larger than the cutoff.
In the post-processing noise-mitigation procedure, when compressing we choose some maximum bond dimension ($\chi_{max}$), and only the largest $\chi_{max}$ singular values are retained. Any singular values beyond this bond dimension are discarded. The computational cost of this type of MPO compression scales as $\chi^3_{max} N$. This method is often used when dealing with large-scale systems with MPOs of very large bond dimension in order to control the computational resources, specifically to limit the memory and computational cost associated with the MPO. 

\subsection{Informationally complete measurements}
\label{sec:informationally_complete_meas}
A \emph{positive operator-valued measure} (POVM) is defined by a set of positive semi-definite operators $\{\Pi_m\}_m$ that act on the Hilbert space of the quantum system and satisfy the completeness relation $\sum_m\Pi_m = \hat{\mathbb{1}}$. The operators $\Pi_m$ are called the POVM effects and represent the different possible outcomes of the measurement $m$. Each effect is associated with a probability given by $p_m = \text{Tr}[\rho \Pi_m]$, where $\rho$ is the density matrix of the quantum state being measured. 
For a POVM to be IC, the POVM effects must form a basis of the space of linear operators in the corresponding Hilbert space with $\text{dimSpan}(\{\Pi_m\}_m) = 4$. Under this assumption, the quantum state is uniquely determined by the probability distribution of outcomes, i.e.
\begin{equation}
    \label{eq:IC_POVM}
    \text{Tr}[\rho \Pi_m] = \text{Tr}[\rho' \Pi_m] , \forall m \iff  \rho = \rho'.
\end{equation} 
If the measurements satisfy Eq.~\eqref{eq:IC_POVM}, one can construct a dual operator $D_m$ for each $\Pi_m$ connecting the probability distribution to the state via the inverse relation
\begin{equation}
\label{eq:Duals}
    \rho = \sum_m \text{Tr}[\rho \Pi_m] D_m=\sum_m p_m D_m \ , \ \forall \rho
\end{equation}
The density matrix $\rho$ can thus be considered as the average dual over the probability distribution of the outcomes of the POVM. For any observable $O$ we can similarly compute the expectation value with our state, $\text{Tr}[O \rho]$ by computing the average $\bar{O} = \sum_m p_m \text{Tr}[OD_m]$.

To construct an IC POVM of an $N$ qubit system, one can perform measurements individually on each qubit. The full system POVM effects and associated dual operators can then be straightforwardly constructed by taking the tensor product of their single qubit elements. This gives the full density matrix equation 
\begin{equation}
    \rho = \sum_{m_0, ...m_{N - 1}} \text{Tr}[\rho \bigotimes^{N - 1}_{q = 0} \Pi_{m_q}^{[q]} ] \bigotimes_{q = 0}^{N - 1} D_{m_q}^{q}] =\sum_{m_0 ... m_{N - 1}}p(\textbf{m}) \bigotimes_{q = 0}^{N - 1}D_{\textbf{m}}
\end{equation}
where $\Pi^{[q]}_{m_q}$ is the POVM effect for measurement $m_q$ performed on qubit $q$, $D_{\textbf{m}} = \bigotimes_{q = 0}^{N - 1} D_{m_q}^{[q]}$ and $\textbf{m} = (m_0,..., m_{N - 1})$.

In practice, we are restricted to a finite number of measurements that can be performed on the quantum device. In this case we form estimates to our quantum state and observable expectation values, incurring some statistical error.
Suppose we run our circuit S times, where one run with some projective measurement on each qubit is referred to as a \emph{shot} ($s$). Hence, we collect S shots and associated outcomes $\textbf{S} = \{m_s\}_s$. The circuit output can then be estimated using the dual operators defined in Eq.~\eqref{eq:Duals} as $\rho_{S} = \frac{1}{S}\sum_{\textbf{m} \in \textbf{S}} D_{\textbf{m}}$. This estimation is exact in the limit of infinitely many circuit executions and its average is unbiased, i.e. $ \rho = \lim_{S \rightarrow \infty} \mathbb{E}[\rho_S]$ . In this case we can define an unbiased estimator for the observable expectation value, 
\begin{equation}
    \label{eq:estimator}
      \bar{O} = \frac{1}{S} \sum_{\textbf{m}} \text{Tr}[D_{\textbf{m}} O],  
\end{equation}
with statistical error resulting from the finite shot number 

\begin{equation}
    \label{eq:var_estimator}
    \Delta\bar{O} = \frac{1}{S} \sqrt{\sum_{\textbf{m} \in \textbf{S}}(\text{Tr}[D_{\mathbf{m}} O] - \bar{O})^2} .
\end{equation}

If we wish to compute the expectation value of an observable not directly on our state $\rho$ but rather on some transformed state $\mathcal{M}(\rho)$, we can easily do so by instead considering the image of the dual operator for each one of the outcomes through the map $\mathcal{M}$ which gives
\begin{equation}
    \bar{O}_{\mathcal{M}} = \frac{1}{S} \sum_{\textbf{m}} \text{Tr}[\mathcal{M}(D_\textbf{m})O] = \frac{1}{S} \sum_{\textbf{m}} \text{Tr}[D_\textbf{m} \mathcal{M}^{\dagger}(O)]. 
\end{equation}

The corresponding statistical error can similarly be evaluated as in Eq.~\eqref{eq:var_estimator} by replacing $O$ with $\mathcal{M}^{\dagger}(O)$.
Importantly, the map $\mathcal{M}$ can be non-physical and can in fact be the inverse of some physical map. 

\subsection{Post-processing of measurement outcomes}
\label{sec:postprocessing_outcomes}
In our experiments, we consider projective single qubit measurements in the eigenbases of the Pauli operators ($X, Y, Z)$ measured according to the probabilities $(p_x, p_y, p_z)$. We use $p_x = p_y = p_z = \frac{1}{3}$ for all qubits but the $n$th qubit (associated with the observable $X_n$ in interest), for which $p_x = 0.8$, $p_y = 0.1$, $p_z = 0.1$. The single qubit POVM effects are
\begin{align}
    \Pi_{z+}  & = p_z|0\rangle \langle0| \quad, \quad \Pi_{z-} = p_z |1\rangle \langle 1|\\ 
    \Pi_{x\pm} & = p_x|\pm\rangle \langle\pm| \quad, \  \Pi_{y\pm} = p_y |\pm i \rangle \langle \pm i |
\end{align}
with respective dual operators
\begin{equation}
    \label{eq:duals_used}
    D_{\alpha \pm} = \frac{1}{2} (\mathbb{1} \pm p^{-1}_{\alpha}P_{\alpha}).
\end{equation}

To collect sufficiently many shots, hardware limitations and simulation time must be taken into account (see Sec.~\ref{sec:circuit_execution}). If we define a circuit setting to be a single instance of the quantum circuit with a certain choice of the measurement basis on each qubit, to align with equations~\eqref{eq:estimator} and~\eqref{eq:var_estimator} we would submit S different settings and collect a single outcome \textbf{m} from each. Since this becomes impractical for large-scale problems, we instead submit a collection of $C$ different circuit settings run $M$ times to collect a total number of $S = CM$ shots. 
For some pairing of shot ($s$) and circuit ($c$), we collect outcome $\textbf{m}(c, s)$, yielding $\xi(c,s) \equiv \text{Tr}[D_{\textbf{m}(c, s)} O]$. The average value for this setting is then $\xi(c) = \frac{1}{M} \sum_s\xi(c,s)$. The unbiased estimator $\bar{O}$ is still computed, like in Eq.~\eqref{eq:estimator} as the average over the total budget of shots 
\begin{equation}
    \label{eq:estimator_settings}
    \bar{O} = \frac{1}{CM} \sum_{c, s} \xi(c, s) = \frac{1}{S} \sum_{\textbf{m}} \xi(\textbf{m}).
\end{equation}
However the statistical error must be rewritten to account for the repeated settings, and now reads
\begin{equation}
    \label{eq:var_estimator_with_settings}
    \Delta \bar{O} = \sqrt{\frac{1}{(CM)^2} \sum_{c,s} [\xi(c,s) - \xi(c)]^2 + \frac{1}{C^2} \sum_c[\xi(c) - \bar{O}]^2}.
\end{equation}
The measured data are processed using tensor networks in the PTM representation (see Sec.~\ref{sec:MPO_PTM}), which provide computationally efficient descriptions of all the objects and the operations involved. Detailed derivations as to how this is achieved are described in Ref.~\cite{filippov2023scalable}(Appendix D).

\begin{figure*}
    \centering
    \includegraphics[width=0.7\columnwidth]{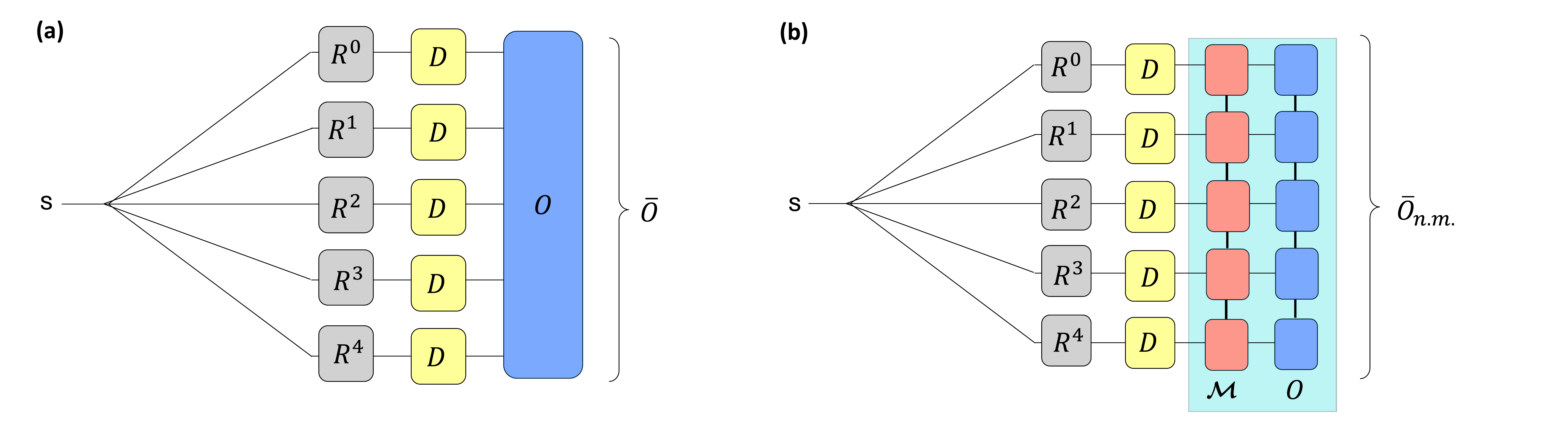}
    \caption[]{
    \textbf{Tensor Network for computing the estimator $\bar{O}$.} $R^{[q]}$ represent the selector matrices with hyperindex $s$ that selects the Dual ($D_{m_q}$) on each qubit for each shot executed on the device. (a) The estimator of the observable $O$, (b) The noise mitigated estimator of the observable modified by the TEM map $\mathcal{M}$.}
    \label{fig:tn_estimator}
\end{figure*}

In a nuthsell, computing the average value of an estimator over the total shots can be implemented via a hyper-indexed tensor network, depicted in Fig.~\ref{fig:tn_estimator}, constructed with the following building blocks: first, on each qubit we have a tensor $D$ representing the set of all single qubit dual operators ($D = \{D_m\}_m$) which can be represented as a matrix in PTM representation. In our case we have the $m = (0, ..., 6)$ dual operators defined in~\eqref{eq:duals_used} giving 
 \begin{equation}
     \label{eq:duals_set}
    D = \frac{1}{\sqrt{2}} \begin{pmatrix}
      1 & p_{x}^{-1} & 0 & 0 \\
      1 & -p_{x}^{-1} & 0 & 0 \\
      1 & 0 & p_{y}^{-1} & 0 \\
      1 & 0 & -p_{y}^{-1} & 0 \\
      1 & 0 & 0 & p_{z}^{-1} \\
      1 & 0 & 0 & -p_{z}^{-1}
    \end{pmatrix}. 
 \end{equation}
We then require corresponding single qubit selector matrices $R^{[q]}$. The selector matrix functions as a way to select the relevant dual operator on each qubit $q$  corresponding to the outcome obtained from each shot $s$. The selector is defined as $R_{sl}^{[q]} = \delta_{m_{q}(s) , l}$ with elements $R_{sl}^{[q]} \in \{0,1\}$ such that $R_{sl}^{[q]} = 1$ if and only if the $q$th element of $\textbf{m}(s)$ is equal to $l$. This means that we can express the quasistate as
\begin{equation}
    \label{eq:tn_quasistate}
    \rho_{\textbf{S}} = \frac{1}{S} \sum_{s = 0}^{S - 1} \bigotimes_{q = 0}^{N - 1} \sum_l R_{sl}^{[q]} D_l
\end{equation}
where the index $s$ is now a hyper-index shared across all qubits containing information about all $S$ outcomes.
Finally, we append the observable ($O$), as an MPO that can be either the original observable (Fig.~\ref{fig:tn_estimator}(a)) or one that has been modified by TEM (Fig.~\ref{fig:tn_estimator}(b)). This gives the full tensor network for the estimator. As described in Sec.~\ref{sec:MPO_PTM}, the action of each of these objects corresponds to a matrix multiplication, thus the contraction for a single shot is equivalent to the calculation of the variable $\xi(s)$. Hence, computing the average value now boils down to simply contracting all shared indices in the tensor network where contraction over the hyper-index corresponds to summing over the $\{\xi(s)\}_s$. 

\subsection{Tensor-network error mitigation (TEM)}
\label{sec:Tensor-network_error_mitigation}

In an ideal scenario, when a simulation is performed on quantum hardware, an initial state evolves to some final state $\rho_{ideal}$. However, due to the presence of noise and imperfections in real quantum hardware, the actual output state $\rho$ will differ from the ideal state due to the influence of some noisy channel $\mathcal{N}$. This is a \emph{completely positive and trace preserving} (CPTP) map that captures the effect of noise during the execution of the quantum circuit which acts on the ideal state $\rho_{ideal}$ resulting in the output $\rho = \mathcal{N}( \rho_{ideal})$. To cancel the effects of noise, we can apply the inverse of this channel to the output to obtain the ideal result. Applying the inverse channel $\mathcal{N}^{-1}$ is not straightforward as the inverse of the CPTP map is itself not necessarily CPTP and therefore is non-physical and cannot be implemented directly on a quantum computer.

\begin{figure*}
    \centering
    \includegraphics[width=\columnwidth]{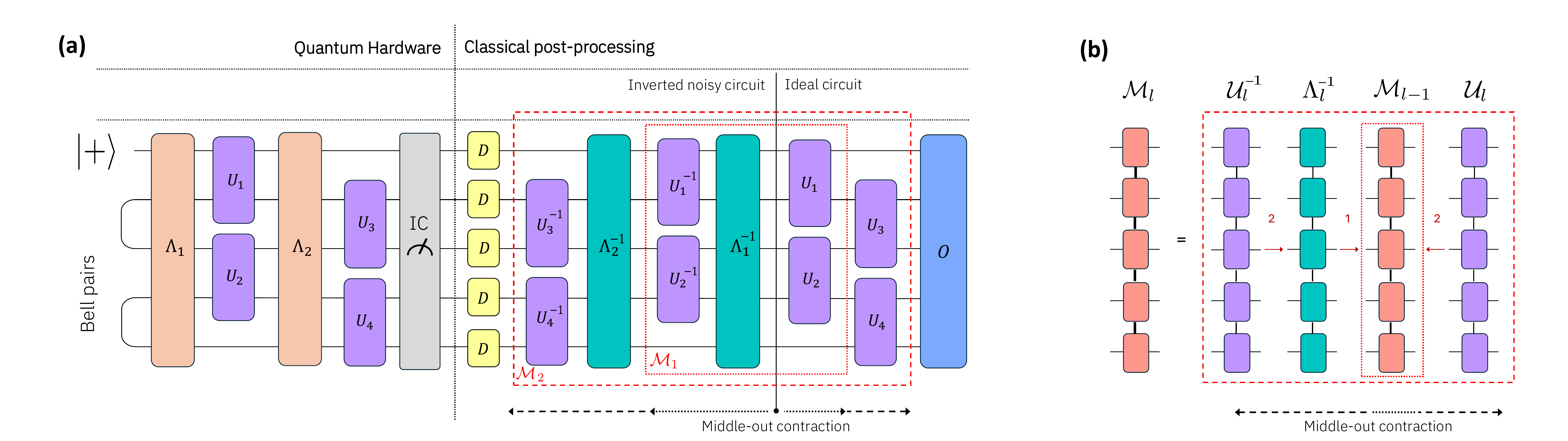}
    \caption[]{
    \textbf{Schematic of tensor-network error mitigation (TEM).}
    (a) Estimation of observables via post-processing of IC measurement data simulated on quantum hardware, with Pauli noise channels $\Lambda_l$ acting before the unitary layers $U_l$. 
    IC measurements yield outcomes to which we assign a dual operator $D$. 
    The TEM mitigation map is applied in post-processing.
    (b) Construction of a  single iteration $\mathcal{M}_l$ of the TEM algorithm as a sequence of contractions and compressions of the layers inside a portion of the full TEM map. 
    Steps 1 and 2 define the order in which we contract and compress within a single iteration.
   }
    \label{fig:tem}
\end{figure*}

Thus, the goal of TEM, as proposed in~Ref.~\cite{filippov2023scalable}, is to shift the cancellation of noise to the post-processing stage using IC measurements, approximating the inverse noise channel with $\mathcal{M} \approx \mathcal{N}^{-1}$ as an MPO, in order to construct the noise mitigated estimator $\bar{O}_{\textbf{n.m.}}$ as 
\begin{equation}
     \label{eq:tem_estimator}
     \bar{O}_{\textbf{n.m.}} = \frac{1}{S} \sum_{\textbf{m}} \text{Tr}[D_\textbf{m} \mathcal{M}^{\dagger}(O)],
\end{equation}
hence bypassing the need for physicality.
Fig. \ref{fig:tem} depicts the full noise mitigation procedure. This procedure is composed of two main parts. The first is performed on the quantum hardware (left) where the IC measurement strategy outlined in Sec.~\ref{sec:informationally_complete_meas} is applied to a circuit with unitary layers $U_l$ each accompanied by a noise channel $\Lambda_l$ acting on the $2$-qubit unitary gates in the layer across all qubits in the register. The $\Lambda_l$ for each layer is assumed to be characterised beforehand via suitable calibrations. In the present case, we will assume that each $\Lambda_l$ is modelled as a sparse Pauli-Lindblad noise channel.
The second part (right) is the classical post-processing portion wherein we append the TEM map as the inverted noisy circuit followed by the ideal noiseless circuit to the quasistate to construct the mitigated result.  
Specifically, the TEM map is defined as 
\begin{equation}
\label{eq:TEM_def_supp}
\mathcal{M} = (\bigcirc_l \mathcal{U}_l) \circ \bigcirc_l (\Lambda_l^{-1} \circ \mathcal{U}_l^{-1} )
\end{equation}
for layers $1,...L$ of the noiseless circuit.

We employ a middle-out contraction strategy to avoid exponential complexity in the number of layers during the multiplication of MPOs. We start from where the inverted noisy circuit meets the ideal circuit and iterate outwards for all $L$ ideal layers of the circuit, availing of the cancellation effects of $\mathcal{U}_l$ and $\mathcal{U}_{l}^{-1}$ such that at each iteration the MPO is close to identity. A single iteration is defined as
\begin{equation}
\label{eq:TEM_iteration}
\mathcal{M}_l = \mathcal{U}_{l} \circ \Lambda_l^{-1} \circ \mathcal{M}_{l - 1} \circ \mathcal{U}_l^{-1}
\end{equation}
that constructs $\mathcal{M}_l$ as an MPO with bond dimension $\chi_l$, where $\mathcal{M}_0 = \mathbb{1}$. This is easily described via tensor networks as depicted in Fig.~\ref{fig:tem}(b).
Specifically, ${\cal U}_l$, ${\cal N}_{l}^{-1}$ and ${\cal U}_{l}^{-1}$ are straightforwardly constructed as MPOs in the PTM representation with bond dimension $4$ (see Sec.~\ref{sec:MPO_PTM}). The sparse Pauli-Lindblad noise defined in Eq.~\eqref{eq:noise_model_def_supp} is a diagonal matrix when expressed in the PTM representation, and its inverse in this form equates to multiplying the generator rates $\lambda_i$ by $-1$. 
Untreated, the growth in the bond dimension for a single iteration $(l-1) \rightarrow l$ of Eq.~\eqref{eq:TEM_iteration} would be 
$\chi_l = 64\chi_{l -1}$. This would be true even in the case of no noise where $\mathcal{M}_l$ is equal to identity. Therefore, in practice, a single iteration is divided into two steps, namely 
\begin{equation}
    (1) \ \mathcal{M}'_l = C_{\chi}(\Lambda_l^{-1} \circ \mathcal{M}_{l - 1})
    \\ \ \ \ \ \ \ \
    (2) \ \mathcal{M}_l = C_{\chi}(\mathcal{U}_l \circ \mathcal{M}'_{l - 1} \circ \mathcal{U}_l^{-1})
\end{equation}
where $C_{\chi}$ indicates compression to some maximum bond dimension $\chi$ (see Sec. \ref{sec:MPO_MPS_compression}). 

Provided the compression errors are reasonably small, the average value of the TEM-modified observable $\mathcal{M}^{\dag}(O)$ in the noisy state $\rho = \mathcal{N}(\rho_{ideal})$ (implemented on hardware) is the same as the noiseless estimation of the original observable $O$. Ref.~\cite{filippov2024scalabilityquantumerrormitigation} clarifies that the main contribution in $\mathcal{M}^{\dag}(O)$ is the rescaled original observable $O$, i.e., $\mathcal{M}^{\dag}(O) \approx c O$ for some $c \geq 1$. In the \textit{typicality scenario}, where the dynamics leads to Pauli branching into a vast number of strings, the Pauli strings in $O$ are damped by the noise by a factor of the same order of magnitude as a randomly chosen Pauli string. If this is the case, then the remaining part of the TEM-modified observable, $\mathcal{M}^{\dag}(O) - c O$, can be essentially neglected as it does not contribute significantly to the average value~\cite{filippov2024scalabilityquantumerrormitigation}. In this case, if $O$ has a low Pauli weight, then the main contribution to the TEM-modified observable has the same low Pauli weight, meaning it can be efficiently estimated with high accuracy via conventional qubit-wise IC measurements. 

\begin{figure}
    \centering
    \includegraphics[width=0.7\columnwidth]{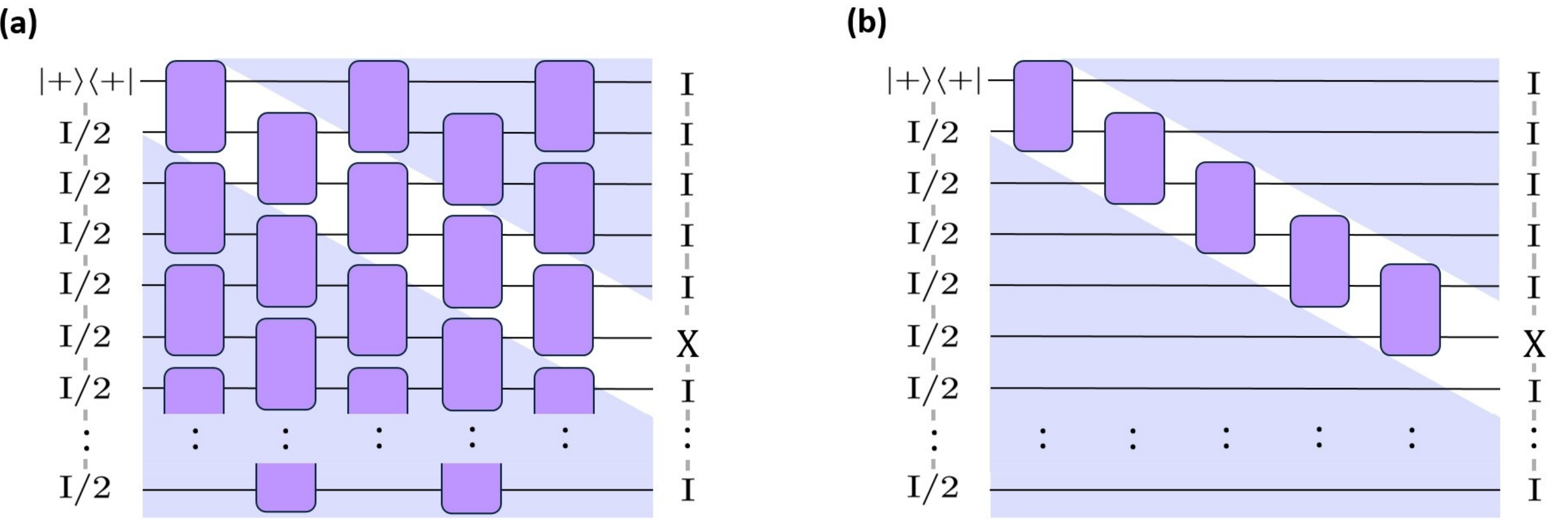}
    \caption[]{
    \textbf{Equivalence of circuits in producing local expectation values}.
    (a) Full circuit.
    (b) Lightcone restricted circuit.
    Left bottom triangle corresponds to the trivial unital dynamics in the Schr\"{o}dinger picture. Right top triangle corresponds to the trivial unital dynamics in the Heisenberg picture. 
   }
    \label{fig:equivalent_circuits}
\end{figure}

The experiment presented in this study goes beyond the typicality scenario as the Pauli strings in the observable do not get scrambled by the evolution into high-weight Pauli strings. In fact, the classical simulations reveal that, in the Heisenberg picture, keeping track of Pauli strings with Pauli weight 2 allows us to reproduce the ideal and noisy signals in the regime of dual unitarity. The signal in non-dual-unitary circuits is simulable in the same way via keeping track of Pauli strings with slightly higher Pauli weight $\lesssim 10$. Non-typicality of the observable leads to high-Pauli-weight components in $M^{\dag}(O) - c O$ that cannot be neglected; however, the estimation of these contributions via the conventional qubit-wise IC measurements leads to a large variance, exponential in the Pauli weight \cite{garcia2021learning}, and calls for more advanced and experimentally demanding measurement techniques \cite{garcia2021learning, glos2022adaptive}. An experimentally friendly way to overcome this problem is to reduce the Pauli weight of the TEM-modified observable via exploiting the degrees of freedom not affecting the noisy and ideal signals. 

For the initial state $\hat{\rho}(0) = \ket{+}_0\bra{+}_0\otimes \hat{\mathbb{1}}^{\otimes (N-1)}/2^{N-1}$, the ideal signal $\langle \hat{X}_n \rangle$ in \textit{any} brickwork unitary circuit is exactly the same as in the simpler circuit depicted in Fig.~\ref{fig:equivalent_circuits}(b). This takes place because the unitary gates do not affect the identity operators propagating from left to right in the Schr\"{o}dinger picture and the identity operators propagating from right to left in the Heisenberg picture. The same statement holds true in the case of Pauli noise as it is unital, i.e., preserves the identity operator. The two circuits are indistinguishable in producing the signal at the $n$th qubit. Therefore, a much simpler circuit in Fig.~\ref{fig:equivalent_circuits}(b) could be used for building the TEM map $\mathcal{M}$.  

If the noise acts locally along the unshaded corridor in Fig.~\ref{fig:equivalent_circuits}(a), then the observable $O = \hat{X}_n$ exhibits a typical behaviour in the Heisenberg picture as its components with different Pauli weight get rescaled by the 2-local noisy maps from the corridor. The resulting typicality leads to a low Pauli weight in the TEM modified observable, thus making its estimation compatible with the IC qubit-wise measurements. In particular, neglecting the terms $\mathcal{M}^{\dag}(O) - c O$, or even including the most significant ones in $\mathcal{M}^{\dag}(O)$, results in an accurate estimate.

If the initial state is $\hat{\rho}(0) = \ket{\Psi (0)} \bra{\Psi (0)}$, $\ket{\Psi(0)} = \ket{+}_0 \otimes \ket{\psi_{\rm Bell}}^{\otimes \lfloor(N-1)/2 \rfloor}$, then the corridor-narrowed location of noisy components (including cross-talk terms overlapping with the corridor) remains valid for dual unitary circuits since the only non-zero contribution to the ideal signal comes from the Pauli string $X_0$ in $\hat{\rho}(0)$. For non-dual-unitary circuits, the Pauli strings contributing to the signal are $X_0$ (leading term) and $Z_{2k}$, $Z_{2k-1}Z_k$ (side terms with amplitudes rapidly decreasing in $k$) as well as negligible $m$th-order contributions $Z_{k_1} \cdots Z_{k_m}$, with the side terms originating from the trajectories along the corridor and a slight deviation from it closer to the end of the evolution in the Heisenberg picture. In this case, the use of the corridor-narrowed noise or of a wider corridor variant serves as a good approximation. We employ the latter approximation in the current experiment to make use of the qubit-wise IC measurements without incurring prohibitive measurement cost.
  
\subsection{Resource effectiveness of TEM and other error mitigation techniques}

Besides TEM, well-established error mitigation strategies operating with the learned noise model are \emph{probabilistic error cancellation} (PEC)~\cite{van2023probabilistic} and \emph{zero-noise extrapolation} (ZNE) with probabilistic error amplification~\cite{kim2023evidence}. Ref.~\cite{filippov2024scalabilityquantumerrormitigation} provides a comparison of the sampling overheads $\Gamma$ and the resulting random errors $\delta$ in mitigated observable estimations for PEC, ZNE, and TEM under realistic sparse Pauli-Lindblad models. 

Two error mitigation techniques $A$ and $B$ result in the same random error $\delta_A = \delta_B$ if the ratio of measurement shots $M_A/M_B$ (used in technique $A$ and technique $B$, respectively) is the same as the ratio of their sampling overheads $\Gamma_A/\Gamma_B$, i.e., $M_A/M_B = \Gamma_A/\Gamma_B$. Therefore, the ratio $\Gamma_A/\Gamma_B$ is the figure of merit for comparing resources needed for implementing technique $A$ rather than technique $B$. If $\Gamma_A/\Gamma_B > 1$, then technique $A$ requires $(\Gamma_A/\Gamma_B)$ times longer execution time than technique $B$ to provide the same accuracy. On the other hand, if the resources are fixed ($M_A = M_B$), then $\delta_A / \delta_B = \sqrt{\Gamma_A/\Gamma_B}$.

The ratio $R = \langle O\rangle_{\rm ideal} / \langle O\rangle_{\rm noisy}$ quantifies the noise strength and can be viewed as $e^{\#_{\rm avg} / 2}$, with $\#_{\rm avg}$ being the average number of errors happening in the relevant part of the circuit associated with the observable propagation in the Heisenberg picture. In the current experiment, the relevant part of the circuit is in the vicinity of the white corridor depicted in Fig.~\ref{fig:equivalent_circuits}(a). Ref.~\cite{filippov2024scalabilityquantumerrormitigation} derives the sampling overheads for PEC, ZNE, and TEM in terms of $\#_{\rm avg}$ and $R$, namely, ${\Gamma_{\rm PEC}} / {\Gamma_{\rm TEM}} = e^{\#_{\rm avg}} = R^2$ and ${\Gamma_{\rm ZNE}} / {\Gamma_{\rm TEM}} = (1 + 1.795 \#_{\rm avg})^2 = (1 + 3.59 \ln R)^2$. 

For experiments with different numbers of qubits, we use the deepest Clifford circuits to estimate $R = \langle O\rangle_{\rm ideal} / \langle O\rangle_{\rm noisy}$ and compare the sampling overheads for different error mitigation strategies on equal footing. The results are presented in Table~\ref{tab:sampling_overheads}.

\begin{table}
\begin{center}
\begin{tabular}{|c||c|c|c|} 
 \hline
$N_\text{qubits}$ & \makecell{$R$} & \makecell{${\Gamma_{\rm PEC}} / {\Gamma_{\rm TEM}}$} &  \makecell{ ${\Gamma_{\rm ZNE}} / {\Gamma_{\rm TEM}}$} \\ 
 \hline\hline
 51 & 3.1 & 9.6 & 25.6 \\ 
 71 & 7.1 & 50.4 & 64.6 \\
 91 & 22.7 & 515 & 149 \\
 \hline
\end{tabular}
\caption{\textbf{Comparison of sampling overheads for different error mitigation techniques.} 
Derivations are based on the signal damping $R = \langle O\rangle_{\rm ideal} / \langle O\rangle_{\rm noisy}$ in the deepest Clifford circuits.}
\label{tab:sampling_overheads}
\end{center}
\end{table}

\subsection{Impact of noise model discrepancies on mitigation outcomes}
\label{sec:relative_error}
Using classical tensor network simulations (see Methods in the main text and Sec.~\ref{sec:classical_simulations}), we can analyse how accurately the learned noise models capture the experimental noise characteristics. 
This provides us with a simulated noisy signal which we would expect the experiments to match assuming the noise model accurately accounts for all noise on the quantum hardware. 
In Sec.~\ref{sec:additional_circuits} we discuss possible reasons why the noise model may differ from the actual noise acting on a single circuit layer. 
While this discrepancy is typically small, deep circuits with many layers can accumulate a non-negligible difference between the simulated outcomes and those obtained from the quantum hardware. 
As is true for any mitigation method that relies on noise characterisation, this mismatch leads to a bias in the mitigated results.

\begin{figure*}[b]
    \centering
    \includegraphics[width=\columnwidth]{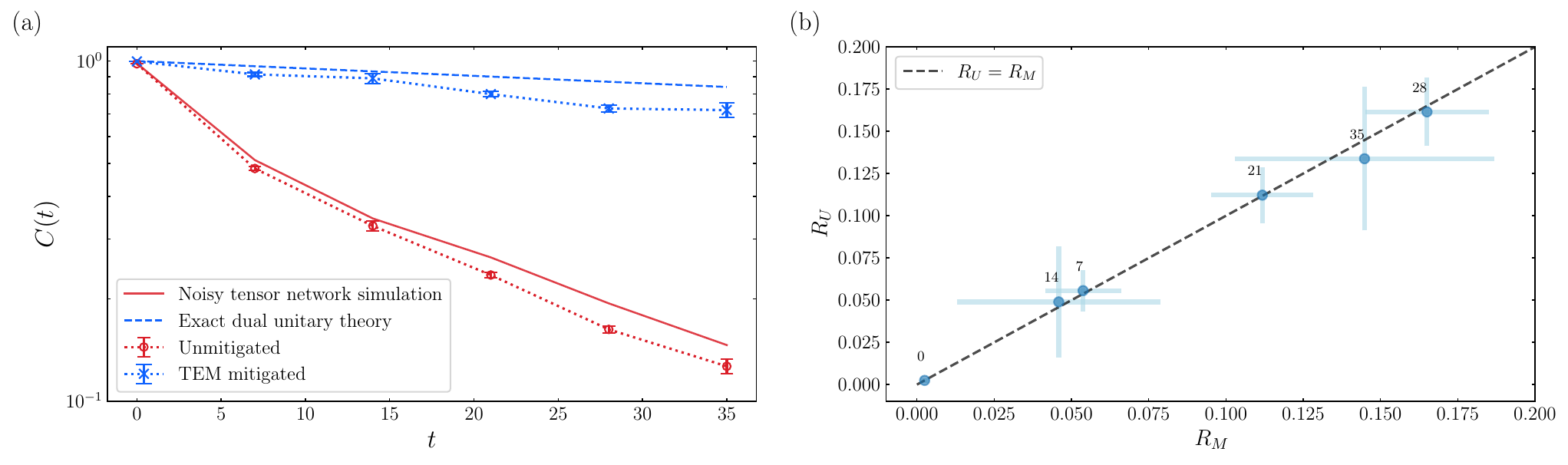}
    \caption[]{\textbf{Impact of noise model discrepancies on mitigated results.}  
    Data shows the autocorrelator for the 71-qubit experiment at the dual unitary point with $h = 0.05$. (a) Unmitigated (mitigated) results compared against classical simulations (exact theory). (b) Correlation in relative errors present in (a) for the unmitigated ($R_U$) and mitigated ($R_M$) case.}
    \label{fig:relative_Error}
\end{figure*}

We study the effect of noise model discrepancies on the error mitigation by considering the relative error between the unmitigated (mitigated) outcomes and the simulated (exact theoretical) results for the experiments at the dual unitary point. 
Here, we exemplify this effect for the $h = 0.05$ curve of the $71$-qubit experiment. 
Let us define the relative error as 
\begin{equation}
    \label{eq:relerror}
    R_{t_{\text{exp}}} = \frac{ | C_{\text{exp}}(t) - C_{\text{ref}}(t) |}{|C_{\text{ref}}(t)|}
\end{equation}
where $R_{t_{\text{exp}}}$ is the relative error in the auto correlator $C_{\text{exp}}(t)$ at time step $t$ and  $C_{\text{ref}}(t)$ is the corresponding reference value. 
Specifically, let $R_U$ denote the relative error in the unmitigated points with reference value obtained from tensor network simulations and let $R_M$ denote the relative error in the mitigated points with the reference value as the exact theoretical result. 
The experimental results for this point are shown in Fig.~\ref{fig:relative_Error}(a) alongside noisy tensor network simulations.
We observe that the deviations of the mitigated curve align closely with those of the unmitigated experiment. 
To examine this effect more closely, we compare the relative errors $R_U$ and $R_M$ in Fig.~\ref{fig:relative_Error}(b).
We find that these errors are consistent, with all data points falling within error bars of the $R_U = R_M$ line. 
We thus conclude that the relative error post-mitigation is consistent with the relative error between the noisy experiment and the expected noisy outcomes. 
This demonstrates that TEM performs as well as can be expected within the margin of error relative to the accuracy of the noise models used.
We note that the same trends discussed here for the $h = 0.05$ curve of the $71$-qubit experiment are also present for other datasets taken at the dual unitary point. 

\section{Classical simulations}
\label{sec:classical_simulations}
\subsection{The Schr\"{o}dinger picture}
A unitary evolution of an $N$-qubit pure state in the Schr\"{o}dinger picture, $\ket{\psi(t)} = {U(t)} \ket{\psi(0)}$, is simulated classically via the \emph{matrix-product-state} (MPS) representation of $\ket{\psi}$ with physical dimension $2$ and some maximum bond dimension $\chi$. The initial state $\ket{\psi(0)}$, composed of Bell pairs in our case, has bond dimension $2$. Any layer of single-qubit gates does not change the bond dimension of the MPS, whereas a layer of ECR gates increases the bond dimension by a factor of $2$. If the resulting bond dimension exceeds the predefined maximum bond dimension $\chi$, the MPS is compressed~\cite{SCHOLLWOCK201196}. The final estimation $\bra{\psi(t)} \hat{X}_n \ket{\psi(t)}$ is the result of a conventional tensor-network contraction~\cite{SCHOLLWOCK201196}.  

\subsection{The Heisenberg picture}
In the Heisenberg picture, we start from the observable ${O}={X}_n$ at the end of the circuit and, step by step, evolve it by ${\cal U}^{\dag} = {U}^{\dag} \bullet {U}$ corresponding to the unitary layer ${U}$, proceeding backwards in time. In the PTM representation, the observable takes the form of an MPS with physical dimension $4$ and the superoperator ${\cal U}^{\dag}$ takes the form of an MPO with physical dimension $4$. The evolution therefore reduces to the sequential application of MPOs to MPS and compression of the resulting MPS if the bond dimension exceeds the predefined maximum value~\cite{dmrg_2007}. Finally, the target expectation value is the overlap of two MPSs: one is the evolved operator in the Heisenberg picture, and the other is the PTM representation of the initial density operator $\ket{\psi(0)}\bra{\psi(0)}$.

\subsection{Convergence of tensor-network classical simulations}
\label{sec:convergence_sim}
\begin{figure}
    \includegraphics{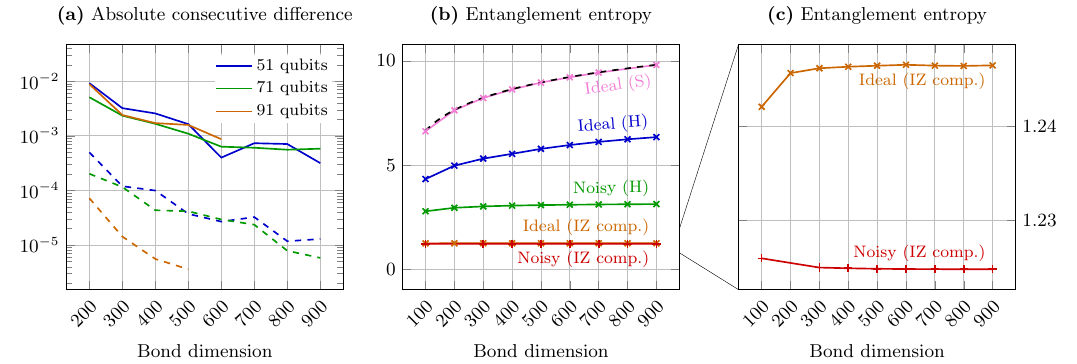}
    \caption{%
        \textbf{Convergence analysis of tensor-network simulations.} 
        (a)~Maximum of the absolute difference, over all values of \(b\) and \(h\), between Heisenberg-picture simulations of the expectation value $\langle \hat{X}_n \rangle$ obtained with bond dimension \(\chi\) and \(\chi-100\).
        Solid lines represent ideal simulations, while dashed lines represent noisy ones.
        (b)~Maximum entanglement entropy, over all values of \(b\) and \(h\), of the final MPS in the ideal and noisy simulations of the 51-qubit circuit (H: Heisenberg picture, S: Schrödinger picture).
        The dashed line indicates the upper bound set by the bond dimension.
        (c)~Focus on the entanglement entropy of the final states after the projection on \(I\) and \(Z\) components.
    }
    \label{fig:tensor_network_simulations_convergence}
\end{figure}

We assess the reliability of the approximate classical simulations by running each tensor-network simulation with a progressively increasing bond dimension \(\chi\) of the MPS representing either the pure state or the observable. 
As \(\chi\) is increased, starting from \(100\), in the Heisenberg-picture simulations we observe that the difference between the expectation values from subsequent runs decreases towards zero, see Figure~\ref{fig:tensor_network_simulations_convergence}(a).
We reached a bond dimension of 900 in the 51- and 71-qubit circuits, while in the 91-qubit case we stopped at 600 due to the expensive memory requirements. 
Remarkably, ideal simulations in the Heisenberg picture approximately follow the same curve, showing that the absolute difference between the result with bond dimension \(\chi\) and \(\chi+100\) does not change more than \(10^{-2}\) for \(100\leq\chi\leq500\) and no more than \(10^{-3}\) for \(500\leq\chi\leq900\).

\begin{figure}
    \centering
    \includegraphics{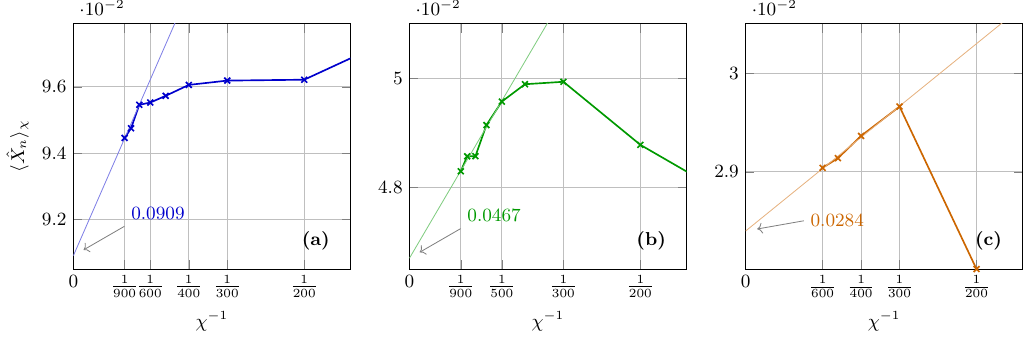}
    \caption{%
    \textbf{Estimated absolute error in tensor-network simulations.} Results of tensor-network simulations in the Heisenberg picture in terms of the inverse of the bond dimension used.
    Illustrative examples for circuits with \(h=\num{0.15}\), \(b=\tfrac{\pi}{4}+\num{0.15}\) at different circuit sizes: \(51\)~(a), \(71\)~(b), and \(91\)~(c) qubits.
    The estimated absolute error is given by Eq.~\eqref{eq:estimated_absolute_error}.
    }
    \label{fig:expected_error}
\end{figure}

As another metric for convergence of the tensor-network simulations, we use the estimated absolute error~\cite[Supplementary Material]{begusic:simulations_utility_qc} defined by
\begin{equation} \label{eq:estimated_absolute_error}
\estabserr_\chi=\abs{\expect{\op{X}_n}_{\chi} - \expect{\op{X}_n}_{\chi \to\infty}},
\end{equation}
where $\expect{\op{X}_n}_{\chi}$ is the average value of the observable in tensor-network simulations with bond dimension $\chi$, and $\expect{\op{X}_n}_{\chi\to\infty}$ is obtained by a linear extrapolation of the data as a function of \(1/\chi\).
More precisely, in each circuit we identify a specific bond dimension \(\bar\chi\) such that \(\expect{\op{X}_n}\) increases approximately linearly as a function of \(\chi^{-1}\) for all \(\chi\geq\bar\chi\). 
For the circuits under study, we obtain $\estabserr_\chi\sim 10^{-2}$ for $100 \leq \chi \leq 400$ and $\estabserr_\chi\sim 10^{-3}$ for $500 \leq \chi \leq 900$.
We illustrate the approach for some of the circuits under study in Fig.~\ref{fig:expected_error}.

We also monitor how the entanglement entropy of the MPS changes as we increase the bond dimensions.
The entanglement entropy is defined as
\begin{equation}
    -\sum_{i=1}^{N}\lambda_i\log_2\lambda_i
\end{equation}
where \(\{\lambda_i\}_{i=1}^{N}\) are the squares of the singular values of the MPS on a certain bond.
More specifically, we consider the maximum of this quantity over all the bonds of the MPS.
The entanglement entropy is bounded from above by \(\log_2\chi\) for an MPS with bond dimensions that does not exceed \(\chi\).
Hence, we can safely say that the truncated bond dimension does not significantly affect the results if we see a plateau well below this value.

In Figure~\ref{fig:tensor_network_simulations_convergence}(b) we show the maximum of the entanglement entropy among all values of \(b\) and \(h\) for the 51-qubit circuit.
The entropy for the 71- and 91-qubit circuits follows the same trend.
We can see that the simulations in the Schrödinger picture achieve the worst possible scaling, hitting the upper bound given by the bond dimension. 
This is due to the highly-entangling nature of the circuit.
The Heisenberg-picture simulations, instead, show an overall lower entanglement entropy.
The particular choice of the observable has a great role in determining the efficiency of the Heisenberg-picture simulations.
In the \(b=\frac{\pi}{4}\), \(h=0\) case we have a Clifford circuit, in which case the observable is mapped to a Pauli-weight-1 string in the end (the string \(ZIII\dotsm\) if the computer is initialised in the state $\ket{0}^{\otimes N}$), and the entropy is zero.
Even when we deviate from these values, by changing \(b\) or \(h\) or by adding noise, the distribution of the coefficients of each component of the observable remains in all cases peaked around the dominant contribution \(ZIII\dotsm\).
During its evolution, the observable picks up other terms with Pauli-weight greater than 1, but their contribution is still not relevant enough to increase the entropy beyond what we can manage with a moderately-sized MPS.

Still, the entanglement entropy of both noisy and ideal simulations in the Heisenberg picture grows logarithmically with the bond dimension. 
This does not contradict the convergence shown in Figure~\ref{fig:tensor_network_simulations_convergence}(a) because we are computing the expectation value of a specific observable in a specific initial state, namely, \( (\ket{0}\bra{0})^{\otimes N}\).
This initial state, in the PTM representation, can be written only with \(I\) and \(Z\) components on each qubit, so it captures only part of the evolved observable.
To take this into account, we recompute the singular values of the final observable MPS after projecting it onto these two components (thus eliminating the contributions containing \(X\) or \(Y\) components, which are orthogonal to the initial state).
In this case not only is the entanglement entropy much lower than in the previous cases, but it is also basically constant for all considered bond dimensions in the range from 100 to 900, which substantiates the convergence analysis.

\begin{figure}
    \centering
    \includegraphics{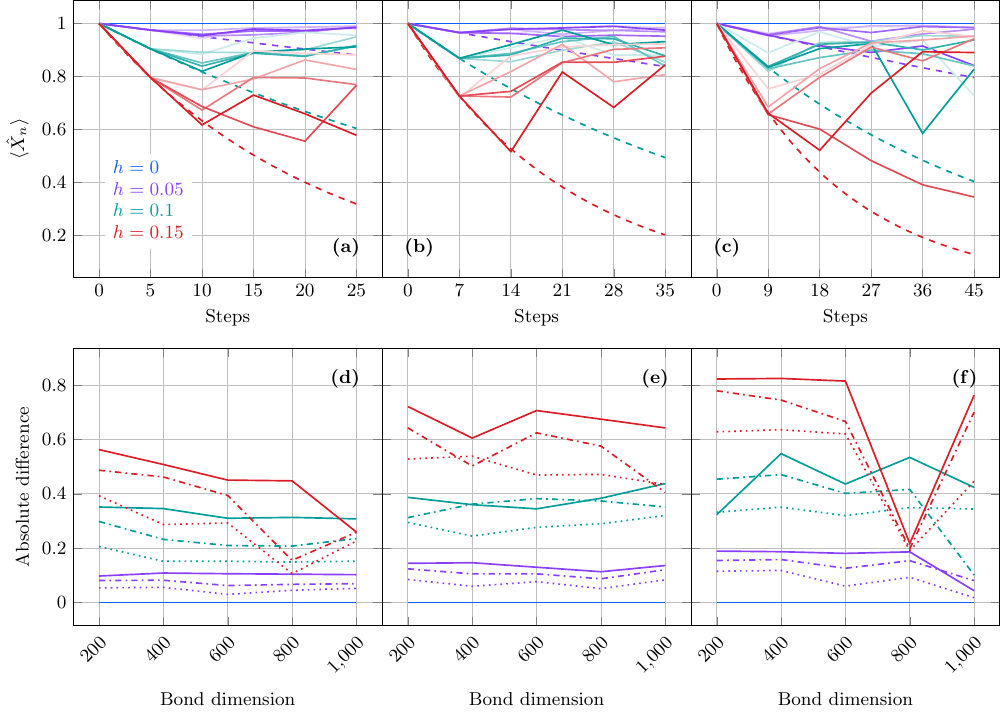}
    \caption{%
    \textbf{Convergence of Schrödinger-picture simulations.}
    (a-c) Simulation of the circuits at the dual-unitary point in the Schrödinger picture, together with the exact theoretical values (dashed) for 51~(a), 71~(b), and 91~(c) qubits.
    The opacity of the curves increases together with the bond dimension, from 200 to 1000.
    (d-f) Absolute differences between simulated and exact values for the last three steps (e.g.~15, 20 and 25 for the 51-qubit circuits; dotted, dash-dotted and solid curves respectively).
    }
    \label{fig:convergence_schroedinger_picture_simulations}
\end{figure}

\subsection{Classical resources}
All tensor-network post-processing and simulation methods were implemented in Python and Julia using a combination of the Quimb~\cite{gray2018quimb} and ITensor~\cite{ITensor} libraries. Simulations were run on the Karolina, Leonardo HPC clusters and Microsoft Azure cloud virtual machines.

\begin{figure}
\includegraphics{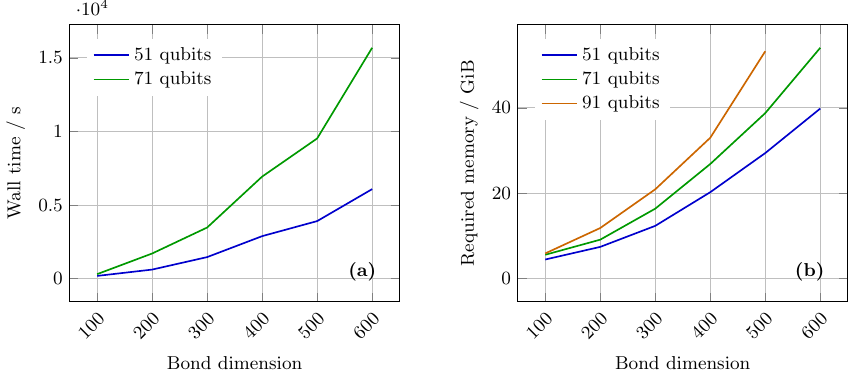}
\caption{%
    \textbf{Time and memory requirements for Heisenberg-picture simulations.}
    (a)~Elapsed time in the simulation of the \(b=\tfrac{\pi}{4}+\num{0.15}\) and \(h=\num{0.15}\) ideal circuit, with 16 CPUs and \qty{72}{\gibi\byte} of memory on the HPC cluster Leonardo, with different bond dimensions.
    (b)~Total memory required for the same simulation (without constraints).
}
\label{fig:classical_resources}
\end{figure}

\begingroup
\newcommand{\schr}{_\textnormal{S}}
\newcommand{\heis}{_\textnormal{H}}
The most expensive part of the simulation is the MPO-MPS multiplication at each layer of the circuit, with time complexity \(O(nd(mm')^3)\) where \(n\) is the number of sites (i.e.~qubits) in the tensor networks, \(d\) is the local dimension and \(m\) and \(m'\) are the bond dimensions of the MPS and the MPO.
In the Schrödinger-picture simulations, the local dimension is 2 and the MPO encoding the 2-qubit layer has bond dimension 2, whereas in the Heisenberg-picture simulations these numbers become 4 and 4.
This means that a Schrödinger-picture with MPS bond dimension \(m\schr\) and a Heisenberg-picture simulation with \(m\heis\) have the same time complexity if \(m\schr=2^\frac43m\heis\).
For example, \(m\schr=2000\) is roughly equivalent to \(m\heis=800\).
Since the kicked-Ising circuit creates less entanglement on the observable side than on the state side, the Schr\"{o}dinger-picture simulations with bond dimension \(m\schr=2000\) are still very far from converging, while the Heisenberg-picture simulations with bond dimension \(m\heis=800\) are sufficient to obtain reliable results.

Figure~\ref{fig:classical_resources} shows how much wall-clock time the ideal (noiseless) evolution of the observable in the Heisenberg picture---for a specific choice of \(b\) and \(h\) associated to one of the most expensive simulations---took with fixed number of CPU cores and amount of memory, as well as the overall memory required in an unconstrained setting (no wall-clock time data is available for the 91-qubit circuit).
As the plot shows, a 91-qubit simulation with bond dimension 500 already requires more than \qty{50}{\gibi\byte}; increasing the bond dimension to 600 would bring the required memory to roughly \qty{90}{\gibi\byte}.
We note that, contrary to the statements above, the time complexity is not proportional to the number of qubits, and the increase does not scale as \(m^3\), but rather as \(m^2\).
The explanation for both of these facts is that, in the Heisenberg picture, the MPS is almost never at ``full capacity''. 
The information spreads linearly along a line from the central qubit, where the observable $ \hat{X}_n $ is supported, with the edges of the circuit being reached only at the very last step in the Heisenberg evolution.
\endgroup

\end{document}